\documentclass[a4paper]{article}
\newcommand{\be} {\begin{equation}}
\newcommand{\ee} {\end{equation}}
\newcommand{\bdm} {\begin{displaymath}}
\newcommand{\edm} {\end{displaymath}}
\newcommand{\bc} {\begin{center}}
\newcommand{\ec} {\end{center}}
\newcommand{\beqa} {\begin{eqnarray}}
\newcommand{\eeqa} {\end{eqnarray}}
\newcommand{\nn} {\nonumber}

\newcommand{\bfig} {\begin{figure}}
\newcommand{\efig} {\end{figure}}
\newcommand{\btab} {\begin{tabular}}
\newcommand{\etab} {\end{tabular}}
\newcommand{\hl} {\hline}
\usepackage{epsfig}
\usepackage{amssymb}
\usepackage{amsmath}
\begin{document}
{\hfill MAN/HEP/2008/2}
\bc
{\bf\Large Scalar Meson Photoproduction}
\ec
\bigskip
\bc
{\large A Donnachie}\\
{\large School of Physics and Astronomy, University of Manchester}\\
{\large Manchester M13 9PL, England}
\ec
\medskip
\bc
{\large Yu S Kalashnikova}\\
{\large ITEP}\\
{\large 117259 Moscow, Russia}
\ec
\medskip
\begin{abstract}
\noindent
The light-quark non-strange scalar mesons $a_0(980)$, $f_0(980)$, 
$f_0(1370)$,
$a_0(1450)$, $f_0(1500)$ and $f_0(1710)$ are of great interest as there 
is no
generally accepted view of their structure which can encompass
$qq\bar{q}\bar{q}$, molecular, $q\bar{q}$ and glueball states in various
combinations. It has been shown previously that the radiative decays of 
the
scalar mesons to $\rho$ and $\omega$ are a good probe of their structure 
and
provide good discrimination among models. Scalar meson photoproduction is
proposed as an alternative to measuring radiative decays and it is shown 
that
it is a feasible proposition.
\end{abstract}

\bigskip

{\large

\section{Introduction}
\label{intro}

The fundamental structure of the light scalar mesons is still a subject 
of controversy. One view is that the $\sigma(485)$, $\kappa(700)$, 
$f_0(980)$ and $a_0(980)$ are molecular or $qq\bar{q}\bar{q}$ states and 
so unrelated to $q\bar{q}$ spectroscopy. Then the $a_0(1450)$ and the 
$K^*_0(1430)$ are regarded as the $u\bar{d}$ and $u\bar{s}$ members of 
the same SU(3) flavour nonet of $1^3P_0$ ground-state $q\bar{q}$ mesons, 
to which the $f_0(1370)$ can be attached as the $(u\bar{u}+d\bar{d})$ 
member \cite{PDG06}. There remain two possibilities for the ninth member 
of the nonet, the $f_0(1500)$ and the $f_0(1710)$. It is frequently 
assumed that this surplus of isoscalar scalars in the 1300 to 1700 MeV 
mass region can be attributed to the presence
of a scalar glueball \cite{AC95}. This assumption is supported by
calculations in quenched lattice gauge theory, which predict a scalar 
glueball in this mass range \cite{MP99,MT05,Chen06}. The three physical 
states are then viewed as mixed $q\bar{q}$ and gluonium states, although 
there is not agreement in detail about the mixing \cite{LW00,CK01}. This is 
the first scenario we consider.

Calculations in unquenched LQCD \cite{LQCD} suggest that the mass of the
lightest glueball could be considerably lower than in the quenched case,  
around 1 GeV, casting doubt on this mixing model and
opening up many other possible interpretations \cite{KZ07}. Further, it 
has been argued that the $f_0(1370)$ may not exist 
\cite{KZ07,Ochs06,MO99}, although this is strongly contested 
\cite{Bugg07,AO08}. Should the $f_0(1370)$
not exist, then the lowest scalar nonet may be taken to comprise the
$a_0(980)$, the $f_0(980)$, the $f_0(1500)$ and the $K_0^*(1430)$. The
$f_0(980)$ and the $f_0(1500)$ are mixed such that the former is close to
a singlet and the latter close to an octet. The lightest scalar glueball 
is a broad object extending from 400 MeV to about 1700 
MeV. This is the second scenario we consider. 

Another possibility is that the $\sigma(485)$, $\kappa(700)$, $f_0(980)$
and $a_0(980)$ comprise the $1^3P_0$ ground-state nonet of $q\bar{q}$,
although their properties are strongly influenced by coupling with a 
low-mass glueball \cite{Nar06}. In this case, in the absence of the 
$f_0(1370)$, the $f_0(1500)$ and $f_0(1710)$ are members of the first 
radial $2^3P_0$ scalar excitation and may also have a significant glueball 
component. This is the third scenario we consider.

Finally it has been suggested \cite{AO08} that the $f_0(1370)$ not only
exists but is a pure octet with strong coupling to $\pi\pi$, making this 
the dominant decay channel in broad agreement with the results of 
\cite{Bugg07}. It is further suggested that the $f_0(1500)$ and 
$f_0(1710)$ (which are the same pole but observed on different Riemann 
sheets) correspond to an unmixed glueball. This is the fourth scenario we 
consider.

Radiative transitions offer a particularly powerful means of probing the
structure of hadrons as the coupling to the charges and spins of the
constituents reveals detailed information about wave functions and can
discriminate among models. If one assumes the scalar mesons to be bound 
$q \bar q$ $^3P_0$ states, the radiative decay proceeds via a quark loop 
and the corresponding matrix element can be estimated in the quark model.
In such framework, both the radiative decay of a vector meson to
a scalar meson, $V \to S\gamma$ \cite{CDK02}, and the radiative decay
of a scalar meson to a vector meson, $S \to V\gamma$ \cite{CDK03} have
been considered previously. The latter appears to be 
the more useful in
practice. It was shown that the radiative decays of the $f_0(1370)$,
$f_0(1500)$ and $f_0(1710)$ are strongly affected by the degree of mixing
between the basis $q\bar{q}$ states and the glueball. It is clear 
\cite{CDK03} that 
the discrimination among the different mixing scenarios provided by the 
radiative decays is strong.

The calculations of \cite{CDK02,CDK03} are relevant for the first 
scenario we consider. They are reviewed briefly in Section \ref{decays} 
and then extended
to calculate the radiative decays of the scalar mesons in the models
appropriate to the second, third and fourth scenarios. This calculation
includes the $a_0(980)$, $f_0(980)$ and $a_0(1450)$ as well as the $f_0(1370)$,
$f_0(1500)$ and $f_0(1710)$.

Photoproduction of the scalar mesons at medium energy provides an 
alternative to direct observation of the radiative decays. It is this 
possibility that we explore here and show that it is viable. The dominant 
mechanism is Reggeised $\rho$ and $\omega$ exchange, both of which are 
well understood in pion photoproduction \cite{GLV97}. The energy  must be 
sufficiently high for the Regge approach to be applicable but not too 
high as the cross section decreases approximately as $s^{-1}$. In 
practice this means approximately 5 to 10 GeV photon energy, which is 
pertinent to Jefferson Laboratory both now and with the proposed upgrade. 
In addition to photoproduction on protons we consider coherent 
photoproduction on $^4$He, encouraged in this by a
recently-approved experiment at Jefferson Laboratory \cite{JLab}. Two
advantages of coherent production are the elimination of background from
baryon resonances, considerably simplifying partial-wave analysis of the
mesonic final state, and the restriction to $\omega$ exchange which is
better understood in photoproduction than is $\rho$ exchange. The
photoproduction model we use is described in Section \ref{model} with
full details of the calculation in the Appendix. Results for the
differential and integrated cross sections on protons and $^4$He in
the narrow-width approximation are also presented in this section. 
 
Mass distributions for specific final states are obtained in Section
\ref{mass}. As it is unlikely that the charged decay modes of the scalars
can be considered because of the very much larger cross sections in
$\pi^+\pi^-$, $K^+K^-$, $2\pi^+2\pi^-$ and $\pi^+\pi^-2\pi^0$ from
vector-meson production we concentrate on all-neutral channels which
automatically exclude any vector-meson contribution. Specifically the
neutral channels are $\pi^0\pi^0$, $\eta^0\eta^0$, $\pi^0\eta^0$ and 
$4\pi^0$. A discussion of the branching fractions of the $f_0(1370)$ is 
included because of the degree of ambiguity associated with this state.

There is a continuum background to the resonance production. A model for
photoproduction of continuum $\pi^0\pi^0$, $\eta^0\eta^0$ and 
$\pi^0\eta^0$ states is presented in Section \ref{continuum}, together 
with examples of the interference between these and appropriate 
resonances. The full details of these calculations are given in the Appendix.

Radiative transitions of scalars can also proceed via intermediate
mesonic loops. Generally, the meson loop mechanism is expected to be
suppressed due to large-$N_C$ considerations. However, in some cases
the meson loop mechanism could be quite relevant, especially in
connection with the $a_0(980)$ and $f_0(980)$ mesons which reside at the
$K \bar K$ threshold, so that the admixture of $K \bar K$ molecule
in their wavefunction is expected to be large. An illustrative example of 
the $f_0(980)$ and $a_0(980)$ mesons photoproduced via a pseudoscalar loop 
mechanism is considered in Section \ref{loops}.

Our conclusions are that suitable combinations of measurements of 
scalar-meson photoproduction can be used to clarify the status of the 
scalars. These are presented in Section \ref{conclusions}.

\section{Radiative Decays}
\label{decays}

The most general structure of the $\gamma S V$
vertex is \cite{KKNHH06}
\be
iF^{\gamma SV}_\mu = g_S\big(q_\mu~(q-k)\cdot\epsilon-
\epsilon_\mu~q\cdot (q-k)\big)
\label{gsv}
\ee
where $\epsilon$ is the photon polarisation vector, $q$ and $k$ are
respectively the 4-momenta of the photon and scalar meson, $m_S$ is the
scalar meson mass, $m_V$ is the mass of vector meson and $(k-q)^2=m_V^2$.

The radiative decay width is \cite{KKNHH06}
\be
\Gamma(S \to \gamma V) = 
g_S^2\frac{m_S^3}{32\pi}\Bigg(1-\frac{m_V^2}{m_S^2}
\Bigg)^3.
\label{width}
\ee

If one assumes the scalar mesons to be bound $q \bar q$ $^3P_0$ states, the
radiative decay proceeds via a quark loop and the corresponding matrix element 
can be estimated in the quark model. The details of the radiative decay 
calculations we use are described in \cite{CDK02,CDK03} and are 
summarised briefly here. Wave functions were assumed to be Gaussian, 
$\exp(-p^2/(2\beta_M^2))$, multiplied by an appropriate polynomial and 
$\beta_M$ was a variational parameter obtained for each state from the 
Hamiltonian
\be
H = \frac{p^2}{m_q}+\sigma r -\frac{4}{3}\frac{\alpha_s}{r}+C.
\label{ham}
\ee
Standard quark-model parameters were used: $\sigma = 0.18$ GeV$^2$,
$\alpha_s = 0.5$, $m_q$ is the quark mass, $0.33$ GeV for $u$ and $d$ 
quarks and $0.45$ GeV for $s$ quarks.

The transition amplitude for the decay at rest of meson $A$, mass $m_A$,
to meson $B$, mass $m_B$, and a photon of three-momentum $\mathbf{p}$ is
\be
\mathbf{M}_{A \to B} =  \mathbf{M}^q_{A \to B} +
\mathbf{M}^{\bar{q}}_{A \to B},
\label{Mdef}
\ee
where $\mathbf{M}^q_{A \to B}$ and $\mathbf{M}^{\bar{q}}_{A \to B}$ describe
the emission from the quark and antiquark respectively. Explicitly, these are
\beqa
{\bf M}^q_{A\to B} &=& {\frac{I_q}{2m_q}}\int d^3k
\big[Tr\{\phi^{\dagger}_B({\bf k}-{\textstyle{\frac{1}{2}}}{\bf p})
\phi_A({\bf k})\big\}(2{\bf k}-{\bf p})\nn\\
&&~~~~-iTr\{\phi_B^{\dagger}({\bf k}-{\textstyle{\frac{1}{2}}}{\bf p})
{\bf \sigma}\phi_A({\bf k})\}\times{\bf p}\big]
\label{Mq}
\eeqa
and
\beqa
{\bf M}^{\bar q}_{A\to B} &=& {\frac{I_{\bar q}}{2m_q}}\int d^3k
\big[Tr\{\phi_A({\bf k})\phi_B^{\dagger}({\bf 
k}+{\textstyle{\frac{1}{2}}}
{\bf p})\}(2{\bf k}+{\bf p})\nn\\
&&~~~~-iTr\{\phi_A({\bf k})
{\bf\sigma}\phi_B^{\dagger}({\bf
k}+{\textstyle{\frac{1}{2}}}{\bf p})\}\times{\bf p}\big]
\label{Mqbar}
\eeqa
where $I_q$ and $I_{\bar q}$ are isospin factors.

The differential decay rate is given by
\be
\frac{d\Gamma}{d~\cos\theta}=p\frac{E_B}{m_A}\alpha I \sum
|\mathbf{M}_{A \to B}|^2,
\ee
where the sum is over final-state polarisations and $I = I^2_q = 
I^2_{\bar{q}}$
is the isospin factor for neutral mesons.

It was shown in \cite{CDK02,CDK03} that the model gives good agreement with
existing data and in \cite{CDK03} that, in general, the uncertainty due 
to the use of Gaussian wave functions is less than $10\%$.

\begin{table}
\bc
\btab{|c|c|c|c|}
\hl
Decay & L & M & H \\
\hl
$f_0(1370) \to \gamma\rho$ &  443 & 1121 & 1540 \\
$f_0(1500) \to \gamma\rho$ & 2519 & 1458 &  476 \\
$f_0(1710) \to \gamma\rho$ &   42 &   94 &  705  \\
\hl
\etab
\caption{Effects of mixing on the radiative decays of the scalars to
$\rho$ \cite{CDK03}. The radiative widths, in keV, are given for the 
three
different mixing scenarios described in the text: light glueball (L),
medium-weight glueball (M) and heavy glueball (H). The radiative decays 
of
the scalars to $\omega$ are $\frac{1}{9}$ of these.}
\ec
\label{radwidths1}
\end{table}

In \cite{CDK03} the radiative decays of the $f_0(1370)$, $f_0(1500)$ and
$f_0(1710)$ were considered assuming that they are mixed states of the
$(u\bar{u}+d\bar{d})$ and $s\bar{s}$ members of the ground-state $1^3P_0$
nonet with a scalar glueball. Three different mixing possibilities have
been proposed \cite{LW00,CK01}: the bare glueball is lighter than the 
bare $n\bar{n}$ state \cite{CK01}; its mass lies between the bare 
$n\bar{n}$ state and the bare $s\bar{s}$ state \cite{CK01}; or it is 
heavier than the bare $s\bar{s}$ state \cite{LW00}. We denote these three 
possibilities by L, M and H respectively. The results from \cite{CDK03} 
for each are given in Table 1.

In principle, an important check on the reliability of these calculations 
and of their application to photoproduction would be provided by the 
radiative decay of the $a_0(1450)$ as this does not have the complication 
of glueball mixing. Again assuming that it is a member of the 
ground-state scalar nonet, its decay width to $\rho\gamma$ is
\be
\Gamma(a_0(1450) \to \rho\gamma) = 298~{\rm keV}
\label{a0decay1}
\ee
and its decay to $\omega\gamma$ is a factor of 9 larger.

\begin{table}
\bc
\btab{|c|c|}
\hl
Decay & Width \\
\hl
$a_0(980) \to \gamma\rho$ & ~14 \\
$f_0(980) \to \gamma\rho$ & ~83 \\
$f_0(1500) \to \gamma\rho$ & 986 \\
\hl
\etab
\caption{Radiative widths in keV of the $a_0(980)$, $f_0(980)$ and 
$f_0(1500)$
to $\rho$ assuming that they are all members of the ground-state nonet, 
and
that the $f_0(980)$ and $f_0(1500)$ are mixed such that the former is a
singlet and the latter is an octet. For the isoscalars
the radiative widths to $\omega$ are $\frac{1}{9}$ of these and that for 
the
$a_0(980)$ is a factor of 9 larger.}
\ec
\label{radwidths2}
\end{table}

In scenario II the $a_0(980)$, $f_0(980)$ and $f_0(1500)$ together with 
the $K_0^*(1430)$ form the ground-state nonet, with the $f_0(980)$ and
$f_0(1500)$ mixed such that the former is close to a singlet and the 
latter close to an octet. The radiative decay of the $f_0(1500)$ can be 
calculated in the same model as before, with the result shown in Table 2. 
For the $a_0(980)$ and $f_0(980)$ we use the results of \cite{KKNHH06}, 
also shown in Table 2, with the $f_0(980)$ width corrected for the 
assumption that the $f_0(980)$ is a singlet.

In scenario III the $a_0(1450)$ is a member of the first radial $2^3P_0$
excitation, as it is in scenario II, together with the $f_0(1500)$,
$f_0(1710)$ and $K_0^*(1430)$. The radiative widths of the $a_0(1450)$ 
and $f_0(1500)$ calculated on the basis of this assumption are given in 
Table 3.
This calculation is much more sensitive to the choice of
parameters than is the ground-state calculation, so the results in Table 
3 are not as reliable as those in Tables 1 and 2.

Finally, scenario IV is analogous to scenario II with the $f_0(1370)$
replacing the $f_0(1500)$ as the octet member of the ground state nonet.
Its decay width to $\rho\gamma$ is
\be
\Gamma(f_0(1370) \to \rho\gamma) = 757~{\rm keV}
\label{octet1370}
\ee

\begin{table}
\bc
\btab{|c|c|}
\hl
Decay & Width \\
\hl
$a_0(1450) \to \gamma\rho$ & ~65 \\
$f_0(1500) \to \gamma\rho$ & 679 \\
\hl
\etab
\caption{Radiative widths in keV of the $a_0(1450)$ and $f_0(1500)$ to 
$\rho$
assuming that they are members of the first radially-excited nonet. The
radiative widths to $\omega$ are a factor larger for the $a_0(1450)$ and
$\frac{1}{9}$ for the $f_0(1500)$.}
\ec
\label{radwidths3}
\end{table}

\section{Scalar Photoproduction}
\label{model}

In this section we develop the formalism for scalar photoproduction and
present the differential and integrated cross sections in the 
narrow-width limit of the scalars for each of the four scenarios 
considered.

\subsection{Cross section formalism}

Let $q$, $p_1$, $k$, $p_2$ be respectively the 4-momenta of the photon,
initial proton, scalar meson and recoil proton. The $\gamma S V$
vertex has the form given in (\ref{gsv}). The $SV\gamma$ coupling, $g_S$,
is obtained from the radiative decay width (\ref{width}) and is assumed 
to be
constant. The $VNN$ vertex is
\be
F^{VNN}_{\nu} = ig_V\gamma_{\nu}-g_T\sigma_{\nu\tau}(p_2-p_1)_{\tau}~.
\label{vnn}
\ee
The $\omega N N$ couplings are rather well defined \cite{RHE87}. We have 
used
$g_V^{\omega} = 15$ and $g_T^\omega =0$ as this gives a good description 
of
$\pi^0$ photoproduction \cite{GLV97}. The $\rho N N$ couplings are not so 
well
defined, with two extremes: strong coupling \cite{RHE87} or weak coupling
\cite{OO78,NBL90,GG94}. We are again guided by pion photoproduction
\cite{GLV97} and choose the strong coupling solution with
$g_V^{\rho} = 3.4$, $g_T^{\rho}= 11$ GeV$^{-1}$.

The vector meson propagator is
\beqa
P^V_{\mu\nu} &=& \frac{1}{m_V^2-t}\Big\{g_{\mu\nu}-\frac{1}{m_V^2}
(p_2-p_1)_\mu(p_2-p_1)_\nu\Big\}\nonumber\\
&=& \frac{1}{m_V^2-t}\Big\{g_{\mu\nu}-\frac{1}{m_V^2}
(q-k)_\mu(q-k)_\nu\Big\}.
\label{prop}
\eeqa

The complete photoproduction amplitude with the vector meson exchange 
mechanism is then
\be
M_\mu(s,t)\epsilon_\mu = \bar{u}(p_2)(A_{\mu\nu}\gamma_\nu+B_\mu)u(p_1)
\epsilon_\mu
\label{amp1}
\ee
where
\be
A_{\mu\nu} = a(g_{\mu\nu}(q\cdot k)-k_\mu q_\nu) =
a(g_{\mu\nu}(q\cdot p)-p_\mu q_\nu),
\label{amunu}
\ee
with
\be
a = \frac{g_S(g_V+2 m_p g_T)}{m_V^2-t},
\ee
and
\be
B_\mu = b(p_{1\mu}(q\cdot k)-k_\mu(q\cdot p_1)) =
b(p_{1\mu}(q\cdot p)-p_\mu(q\cdot p_1)),
\label{bmu}
\ee
with
\be
b = -\frac{2g_Sg_T}{m_V^2-t}.
\ee
For the exchange of a single vector meson we find (see Appendix),
\beqa
|M(s,t)|^2 &=& -\textstyle{\frac{1}{2}}aa^*(s(t-t_1)(t-t_2)+
\textstyle{\frac{1}{2}}st(t-m_S^2)^2) \nonumber\\
&&-\textstyle{\frac{1}{2}}(ab^*+a^*b)m_ps(t-t_1)(t-t_2)\nonumber\\
&&-\textstyle{\frac{1}{8}}bb^*s(4m_p^2-t)(t-t_1)(t-t_2).
\label{msquare}
\eeqa
where $t_1$ and $t_2$ are the kinematical bounaries given by equations 
(\ref{t12a}), and the differential cross section is
\be
\frac{d\sigma}{dt}=\frac{|M(s,t)|^2}{16\pi(s-m_p^2)^2}.
\label{sigma}
\ee

\subsection{Reggeisation}

The standard prescription for Reggeising the Feynman propagators in
(\ref{msquare}), assuming a linear Regge trajectory $\alpha_V(t)= 
\alpha_{V0}+\alpha^\prime_V t$, is to make the replacement
\be
\frac{1}{t-m_V^2} \to \Big(\frac{s}{s_0}\Big)^{\alpha_V(t)-1}
\frac{\pi\alpha^\prime_V}{\sin(\pi\alpha_V(t))}
\frac{-1+e^{-i\pi\alpha_V(t)}}{2}\frac{1}{\Gamma(\alpha_V(t))}.
\label{regge}
\ee
This simple prescription automatically includes the zero observed at
$t \approx -0.5$ GeV$^2$ in both $\rho$ and $\omega$ exchange and 
provides a satisfactory description of the $\rho$ and $\omega$ exchange 
contributions to pion photoproduction \cite{GLV97}. We know that this 
approximation is not precise as there are additional contributions, in 
particular from Regge cuts that are clearly required by finite-energy sum 
rules \cite{Wor72,BDS74,BS78}
and, for $\pi^0$ photoproduction, from the trajectory associated with
$b_1(1235)$ exchange \cite{GLV97,BDS74,BS78}. However the overall effect
of these additional contributions is small, the principal effects being 
to weaken the dip in the cross section and to modify the energy
dependence at large $|t|$. The prescription (\ref{regge}) does not 
require the addition of form factors at either vertex when applied to 
pseudoscalar photoproduction so we adopt the same procedure here. We 
assume non-degenerate $\rho$ and $\omega$ trajectories
\beqa
\alpha_\rho &=& 0.55 + 0.8 t\nonumber\\
\alpha_\omega &=& 0.44 + 0.9 t
\eeqa

For photoproduction on $^4$He we assume that the cross section is given 
by
\be
\frac{d\sigma(\gamma N \to f_0 {\rm He})}{dt} = \frac{d\sigma(\gamma N 
\to
f_0 N)}{dt}\Big(4F_{\rm He}(t)\Big)^2,
\label{helium1}
\ee
where $F_{\rm He}(t)$ is the helium form factor \cite{MS91}
\be
F_{\rm He}(t) \approx e^{9 t}.
\label{helium2}
\ee
The justification for the assumption (\ref{helium1}) is the low level of
nuclear shadowing observed on $^4$He at the energies with which we are
concerned, for both pion and photon total cross sections \cite{GS78}.
Writing
\be
\sigma_{hA} = A_{\rm eff}\sigma_{hN},
\label{shadowing}
\ee
where $h$ can be a pion or a photon, it is found that $A_{\rm eff}
\approx 0.9$ at the energies in which we are interested. Further, the
detailed behaviour of $A_{\rm eff}$ as a function of photon energy and
nucleus is rather well described by a simple vector-dominance model
\cite{GS78}.

\subsection{Narrow-width Cross Sections}
\label{narrow}

\subsubsection{Scenario I}
\label{scen1}

\begin{table}
\bc
\vskip2.5mm
\btab{|c|c|c|c|c|c|c|}
\hl
              & \multicolumn{3}{c|}{proton}
              & \multicolumn{3}{c|}{$^4$He}         \\
\hl
Scalar & L & M & H & L & M & H \\
\hl
$f_0(1370)$ & 27.1 & 68.6 & 94.2 & 0.64 & 1.63 & 2.23 \\
$f_0(1500)$ & 89.9 & 52.1 & 17.0 & 1.55 & 0.90 & 0.29 \\
$f_0(1710)$ &  0.7 &  1.6 & 11.8 & 0.0026 & 0.0058 & 0.043 \\
\hl
\etab
\ec
\caption{Integrated photoproduction cross sections in nanobarns on 
protons
and $^4$He at $E_\gamma = 5$ GeV for the three different mixing 
scenarios:
light glueball (L), medium-weight glueball (M) and heavy gluebal (H).}
\label{sigtot1}
\end{table}

\bfig
\bc
\begin{minipage}{70mm}
\epsfxsize70mm
\epsffile{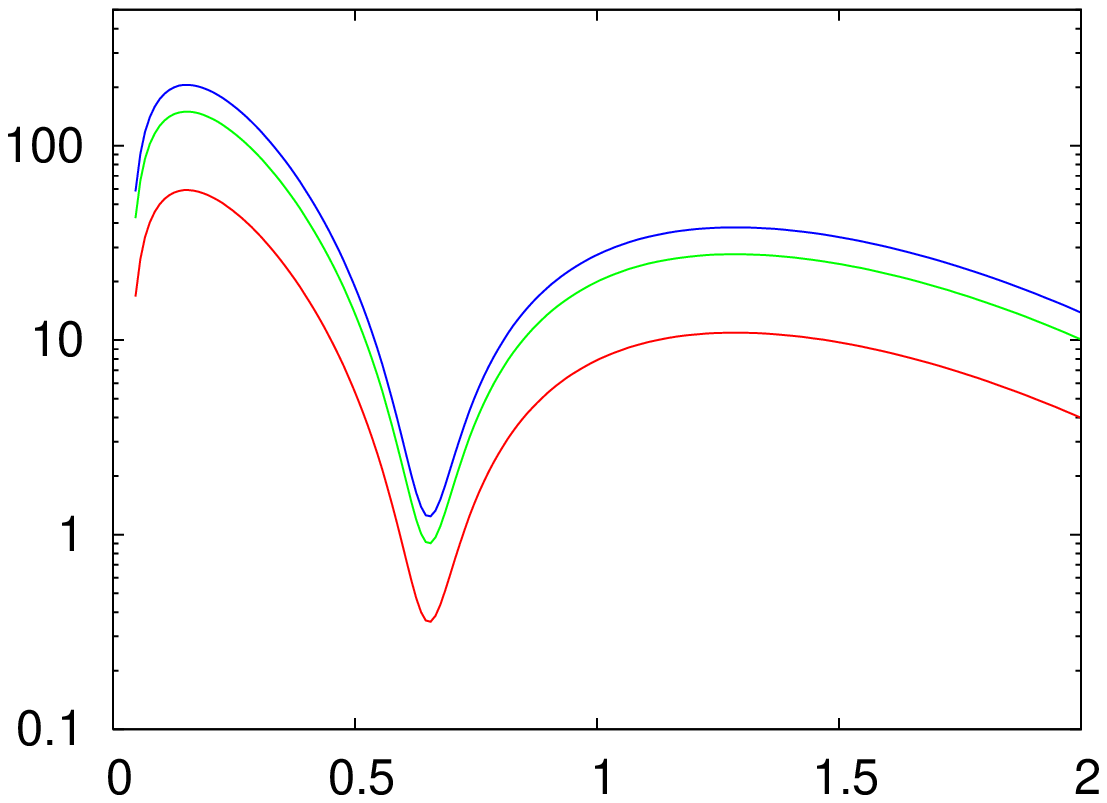}
\begin{picture}(0,0)
\setlength{\unitlength}{1mm}
\put(0,38){$\frac{d\sigma}{dt}$}
\put(40,1){{\small $-t$ (GeV$^2$)}}
\end{picture}
\end{minipage}
\begin{minipage}{70mm}
\epsfxsize70mm
\epsffile{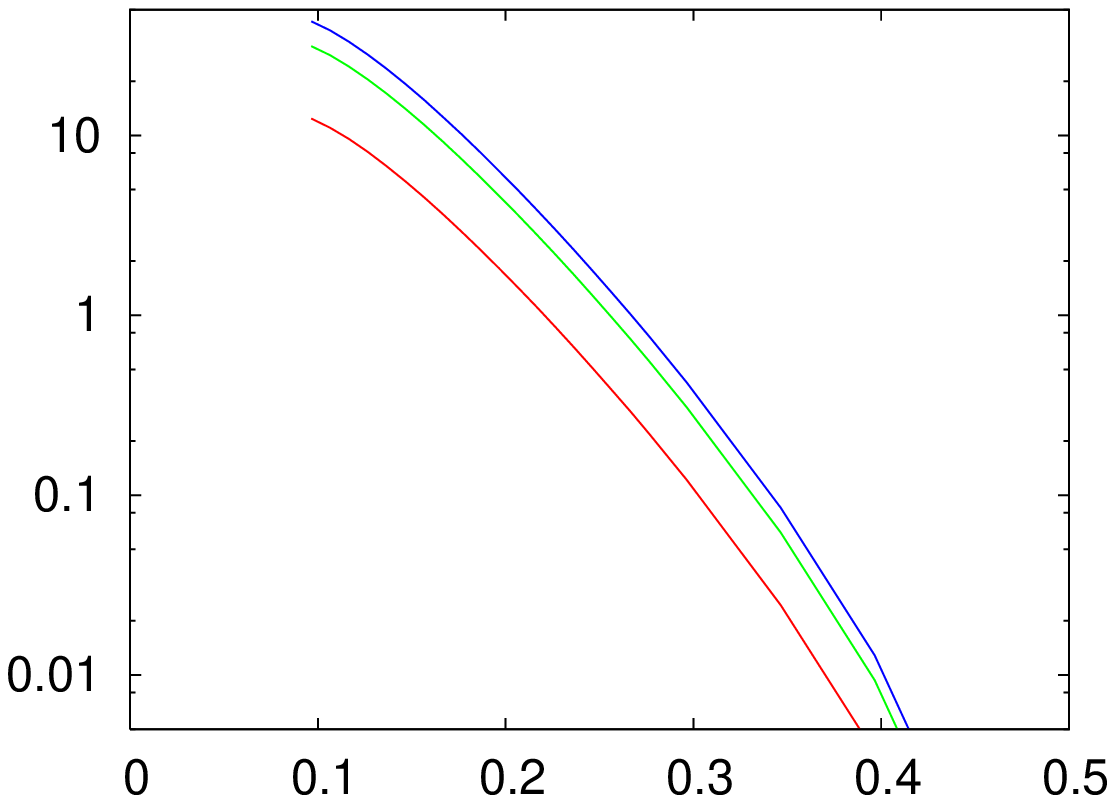}
\begin{picture}(0,0)
\setlength{\unitlength}{1mm}
\put(40,1){{\small $-t$ (GeV$^2$)}}
\end{picture}
\end{minipage}
\vskip 5truemm
\begin{minipage}{70mm}
\epsfxsize70mm
\epsffile{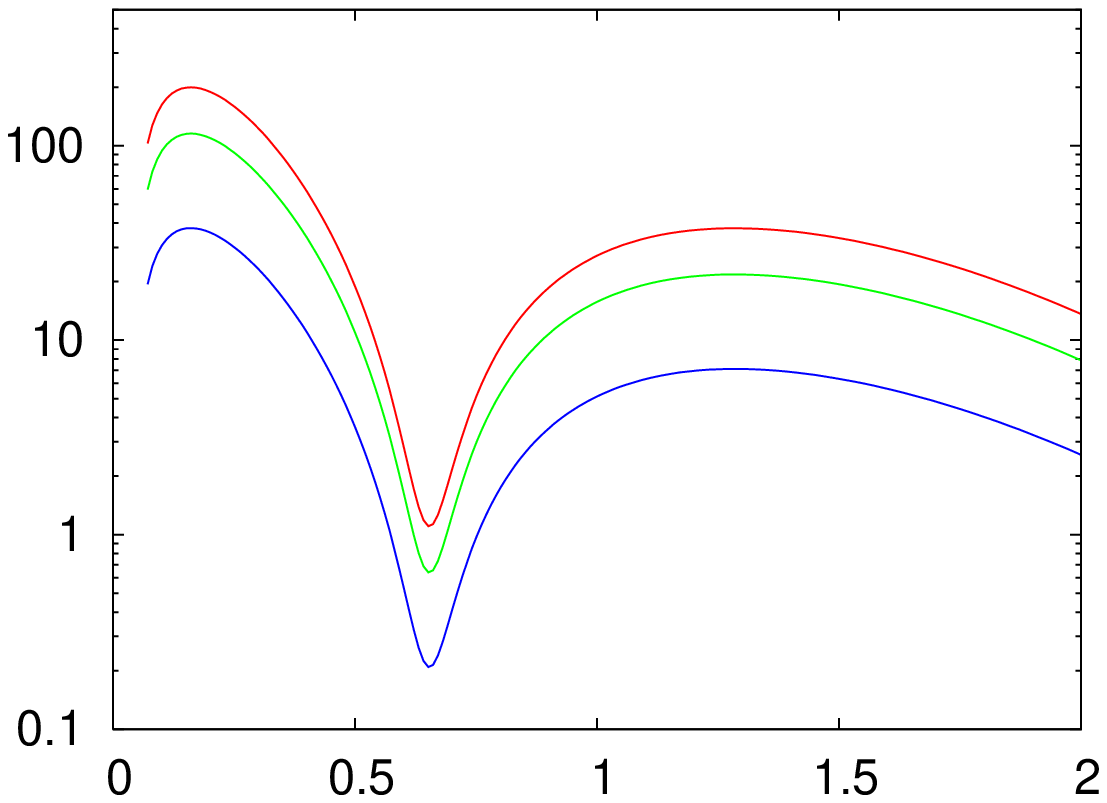}
\begin{picture}(0,0)
\setlength{\unitlength}{1mm}
\put(0,38){$\frac{d\sigma}{dt}$}
\put(40,1){\small{$-t$ (GeV$^2$)}}
\end{picture}
\end{minipage}
\begin{minipage}{70mm}
\epsfxsize70mm
\epsffile{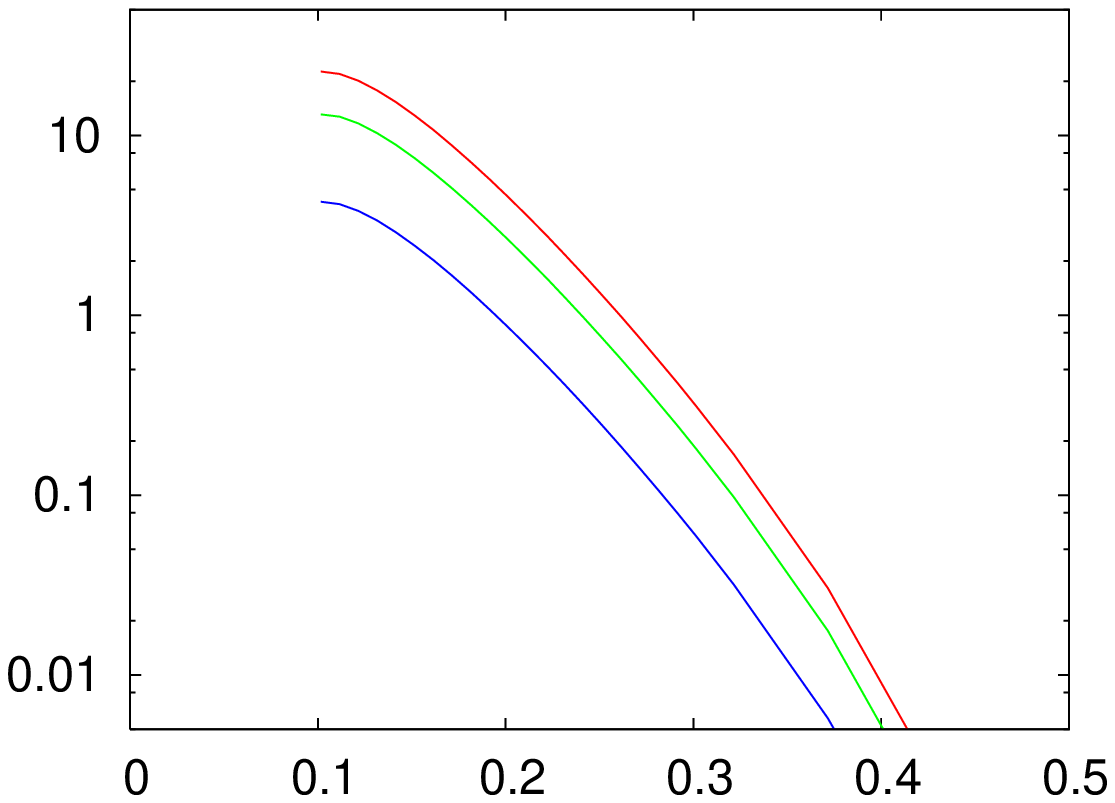}
\begin{picture}(0,0)
\setlength{\unitlength}{1mm}
\put(40,1){{\small $-t$ (GeV$^2$)}}
\end{picture}
\end{minipage}
\vskip 5truemm
\begin{minipage}{70mm}
\epsfxsize70mm
\epsffile{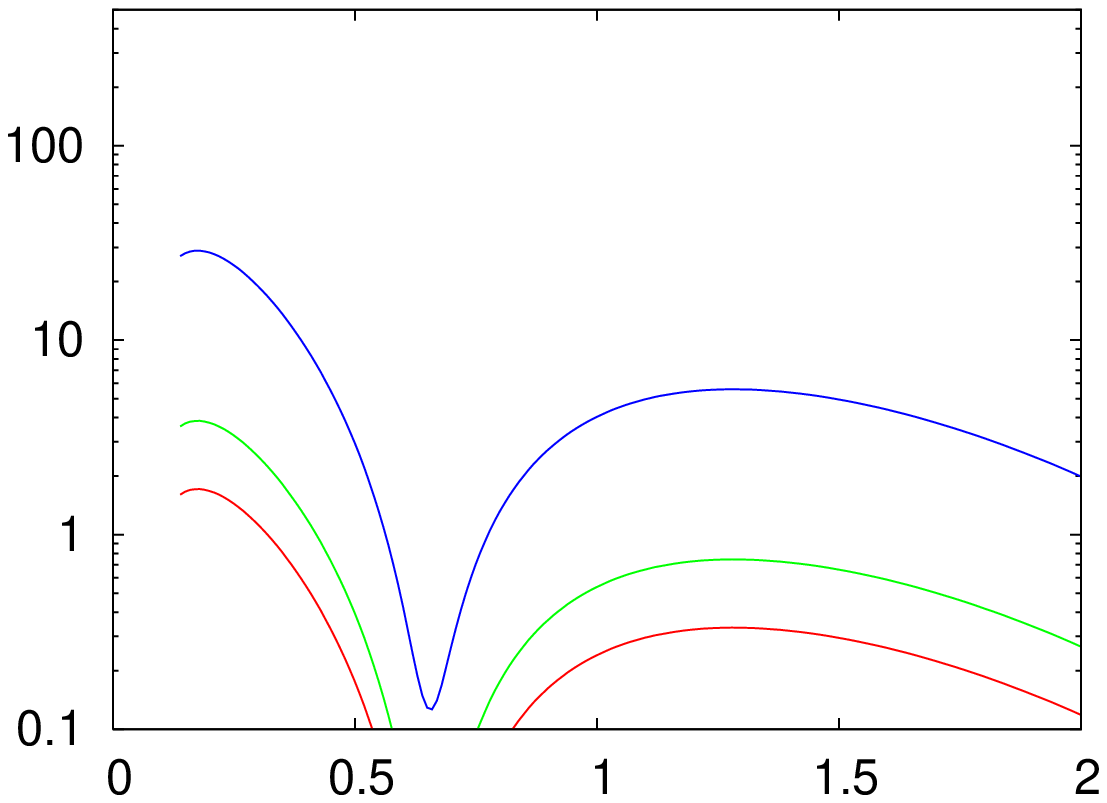}
\begin{picture}(0,0)
\setlength{\unitlength}{1mm}
\put(0,38){$\frac{d\sigma}{dt}$}
\put(40,1){{\small $-t$ (GeV$^2$)}}
\end{picture}
\end{minipage}
\begin{minipage}{70mm}
\epsfxsize70mm
\epsffile{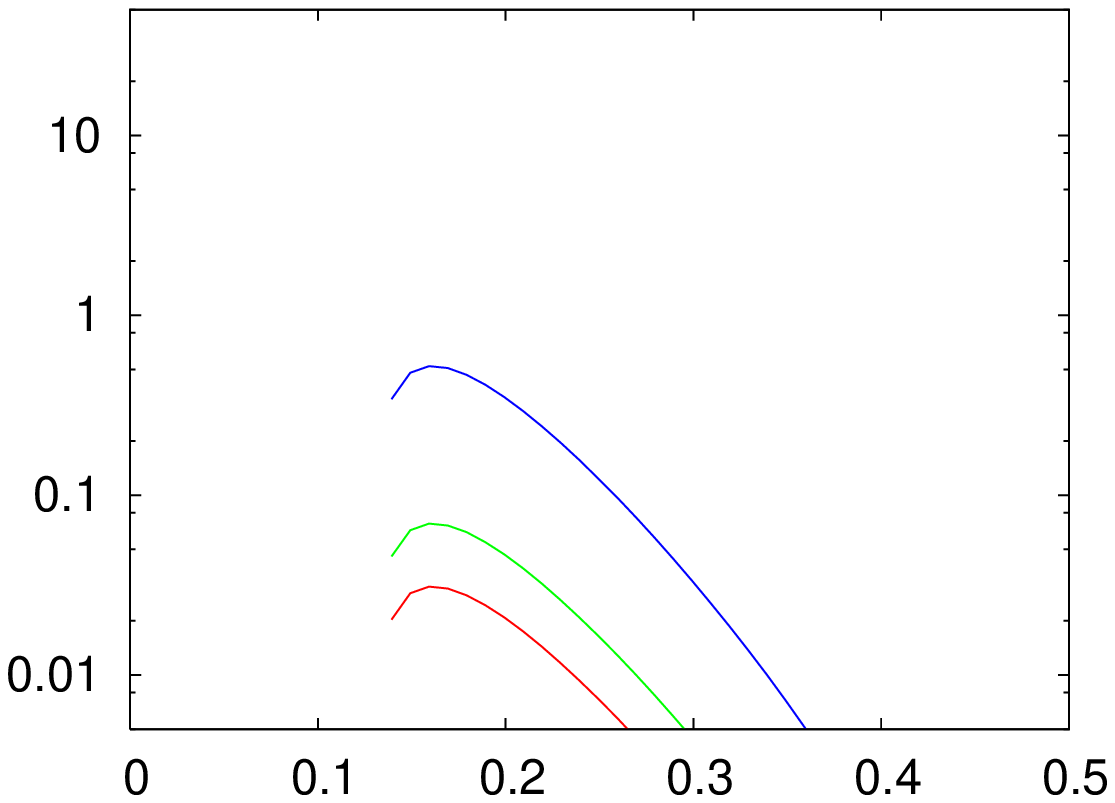}
\begin{picture}(0,0)
\setlength{\unitlength}{1mm}
\put(40,1){{\small $-t$ (GeV$^2$)}}
\end{picture}
\end{minipage}
\caption{Scenario I. Differential photoproduction cross sections on 
protons
(left column) and $^4$He (right column) in nb GeV$^{-2}$ for $f_0(1370)$ 
(top
row), $f_0(1500)$ (middle row) and $f_0(1710)$ (bottom row) at $E_\gamma 
=5$
GeV. The glueball masses are L (red), M (green) and H (blue) in each 
figure.}
\ec
\label{dsigdt1}
\efig

The differential cross sections for photoproduction of the $f_0(1370)$,
$f_0(1500)$ and $f_0(1710)$ on protons and $^4$He at $E_\gamma$ = 5 GeV 
are shown in Figure 1 and the integrated cross sections are given in 
Table 4. In each case results are given for the three possible glueball 
masses: light (L), medium (M) and heavy (H). The cross sections for 
photoproduction on protons decrease with energy at the rate expected from 
(\ref{regge}) so at $E_\gamma = 10$ GeV are about half of those in Table 4. 
However the cross sections for photoproduction on $^4$He do not
decrease, and for the $f_0(1500)$ and $f_0(1710)$ actually increase. This 
is due to the combined effect of the $^4$He form factor enhancing the 
contribution from small $t$ and the maximum of the differential cross 
section on protons moving to smaller $t$ with increasing energy. Note 
that for $^4$He it is necessary to have $|t| \gtrsim 0.1$ GeV$^2$ as this 
is the minimum achievable momentum transfer at which the recoiling 
$\alpha$-particle can be detected \cite{JLab2}.

The cross sections for photoproduction on $^4$He are very much smaller
than those for photoproduction on protons. There are three reasons for
this.
\begin{itemize}
\item
Switching off $\rho$ exchange for photoproduction on protons reduces the 
cross section by a factor of about 16, cancelling the factor 16 from 
coherent production.
\item
The helium form factor suppresses the cross section except at very small 
$t$.
\item
There is the experimental requirement that $|t| \gtrsim 0.1$ GeV$^2$ for 
the recoiling helium to be detected.
\end{itemize} 

\bfig
\begin{minipage}{70mm}
\epsfxsize70mm
\epsffile{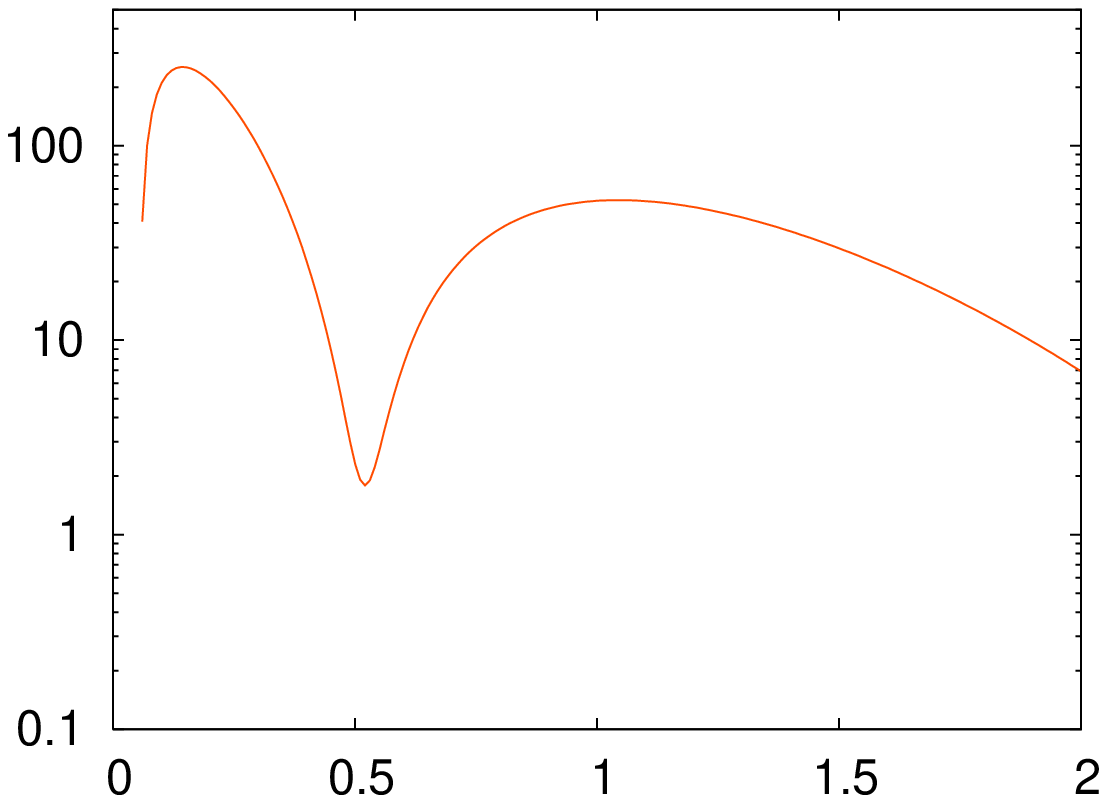}
\begin{picture}(0,0)
\setlength{\unitlength}{1mm}
\put(0,38){$\frac{d\sigma}{dt}$}
\put(55,45){{\small (a)}}
\put(40,1){{\small $-t$ (GeV$^2$)}}
\end{picture}
\end{minipage}
\begin{minipage}{70mm}
\epsfxsize70mm
\epsffile{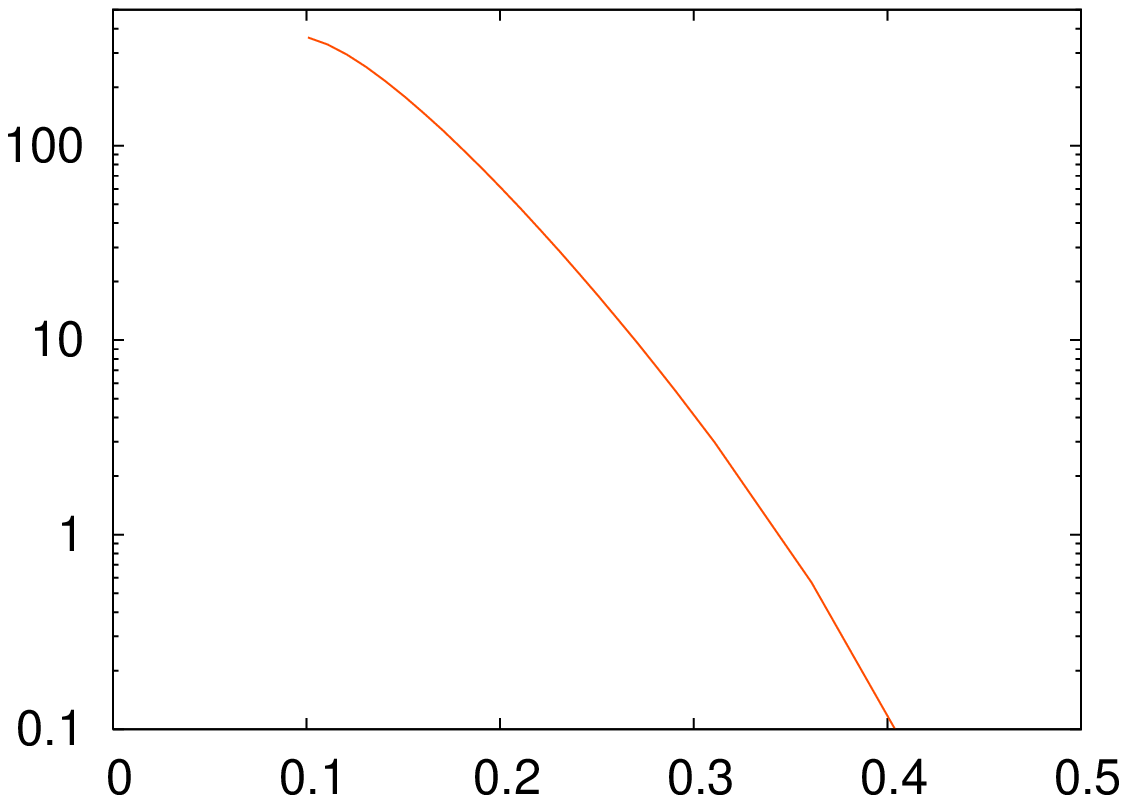}
\begin{picture}(0,0)
\setlength{\unitlength}{1mm}
\put(55,45){{\small (b)}}
\put(40,1){{\small $-t$ (GeV$^2$)}}
\end{picture}
\end{minipage}
\caption{Differential photoproduction cross sections in nb GeV$^{-2}$ for
ground-state $a_0(1450)$ on (a) protons and (b) $^4$He at $E_\gamma = 5$
GeV.}
\label{dsig2}
\efig

Obviously the cross sections for light, medium and heavy glueball masses
reflect directly the radiative decay widths of Table 1 and,
if it were practical, ratios of cross sections
$f_0(1370):f_0(1500):f_0(1710)$ would give an immediate result and
``weigh'' the glueball. The change from L to H is more than a factor of
five, L to M nearly a factor of two and M to H nearly a factor of three.
Coherent production on $^4$He, if practical, would be particularly
important quite apart from the elimination of contributions
from excited baryons. Not only are the trajectories associated with
$\rho$ and $b_1(1235)$ exchange excluded, any Regge cut effects should be
comparable for each scalar, so in the ratios the uncertainty in their
contribution will be minimised. The ratios also remove any ambiguity
associated with form factors and the $\omega N N$ coupling. However, as
we shall see, the situation is not nearly so clear-cut when we come to
consider particular final hadronic states. In particular, once the
standard $\pi\pi$ branching fractions (see Table 7 in Section 4) are 
taken  into account the cross sections for $f_0(1370)$, $f_0(1500)$ 
and $f_0(1710)$ photoproduction on $^4$He in specific channels are 
too small to be practical. However the $f_0(1370)$ would be an 
exception if the $\pi\pi$ branching fraction suggested by Bugg 
\cite{Bugg07} is correct. See the discussion relating to Table 7 
in Section 4.

The integrated photoproduction cross sections for $a_0(1450)$ are $98$ nb 
on protons and $21$ nb on $^4He$. In contrast to the isoscalars, the
cross section for photoproduction of the isovector $a_0(1450)$ on $^4$He
is not strongly suppressed as its dominant radiative decay is to
$\omega\gamma$. The differential cross sections for $a_0(1450)$ 
photoproduction on protons and $^4$He are shown in Figure 2. In this 
scenario the $a_0(980)$ and $f_0(980)$ are not $n\bar{n}$ states and
discussion of them is deferred until Section 5.

\subsubsection{Scenario II}
\label{scen2}

In this scenario the lowest nonet now comprises the $a_0(980)$,
$f_0(980)$ (singlet), $f_0(1500)$ (octet) and the $K_0^*(1430)$. The
$a_0(1450)$ is assigned to the $2^3P_0$ radial excitation. The integrated
cross sections for photoproduction of the $a_0(980)$ and $f_0(980)$ on
protons and $^4$He at $E_\gamma$ = 5 GeV are given in Table 5 and the
differential cross sections in Figure 3. As for the $a_0(1450)$ in
scenario I, the cross section for photoproduction of the isovector
$a_0(980)$ on $^4$He is large. The large photoproduction cross
sections for the $a_0(980)$ and $f_0(980)$ in this scenario are a direct
consequence of their being $n\bar{n}$ states. A corollary is that if the
$a_0(980)$ or $f_0(980)$ are non-$n\bar{n}$ states, as in scenario I,
even a small $n\bar{n}$ admixture will lead to a significant increase in
the cross section. The integrated cross sections for the $f_0(1500)$ and
the $a_0(1450)$ are also given in Table 5 and the differential cross
sections on protons in Figure 3.

\begin{table}[t]
\bc
\btab{|c|c|c|}
\hl
State & proton & $^4$He \\
\hl
$a_0(980)$ & 167.9 & 47.2 \\
\hl
$f_0(980)$ & 91.0 & 3.5 \\
\hl
$f_0(1500)$ & 35.4 & 0.6 \\
\hl
$a_0(1450)$ & 21.4 & 4.7 \\
\hl
\etab
\caption{Scenario II. Integrated photoproduction cross sections in 
nanobarns on protons and $^4$He at $E_\gamma = 5$ GeV for the $a_0(980)$, 
$f_0(980)$, $f_0(1500)$ and $a_0(1450)$ assuming that the isoscalars are 
members of the ground-state nonet, that the $f_0(980)$ is pure singlet and the 
$f_0(1500)$ is pure octet and that the $a_0(1450)$ is the first radial 
excitation.}
\ec
\label{sigtab3}
\end{table}

\subsubsection{Scenario III}
\label{scen3}
As in scenario II, the $a_0(980)$ and $f_0(980)$ are in the $n\bar{n}$
ground-state nonet but the $f_0(980)$ is no longer a singlet. The
integrated cross section and differential cross section for the $a_0(980)$ 
are as in Table 5 and Figure 3, but the results for the $f_0(980)$ in Table 
5 and Figure 3 need to be scaled up by a factor of 1.5. Both the $a_0(1450)$
and $f_0(1500)$ are now members of the first radially-excited nonet. The
results for the $a_0(1450)$ are as in Table 5 and Figure 3(d). The
integrated cross sections for the radially-excited $f_0(1500)$ are 24.7nb 
on protons and 0.4nb on $^4He$. The differential cross section for 
$f_0(1500)$ on protons is shown in Figure 4(a).

\bc
\bfig[t]
\begin{minipage}{70mm}
\epsfxsize70mm
\epsffile{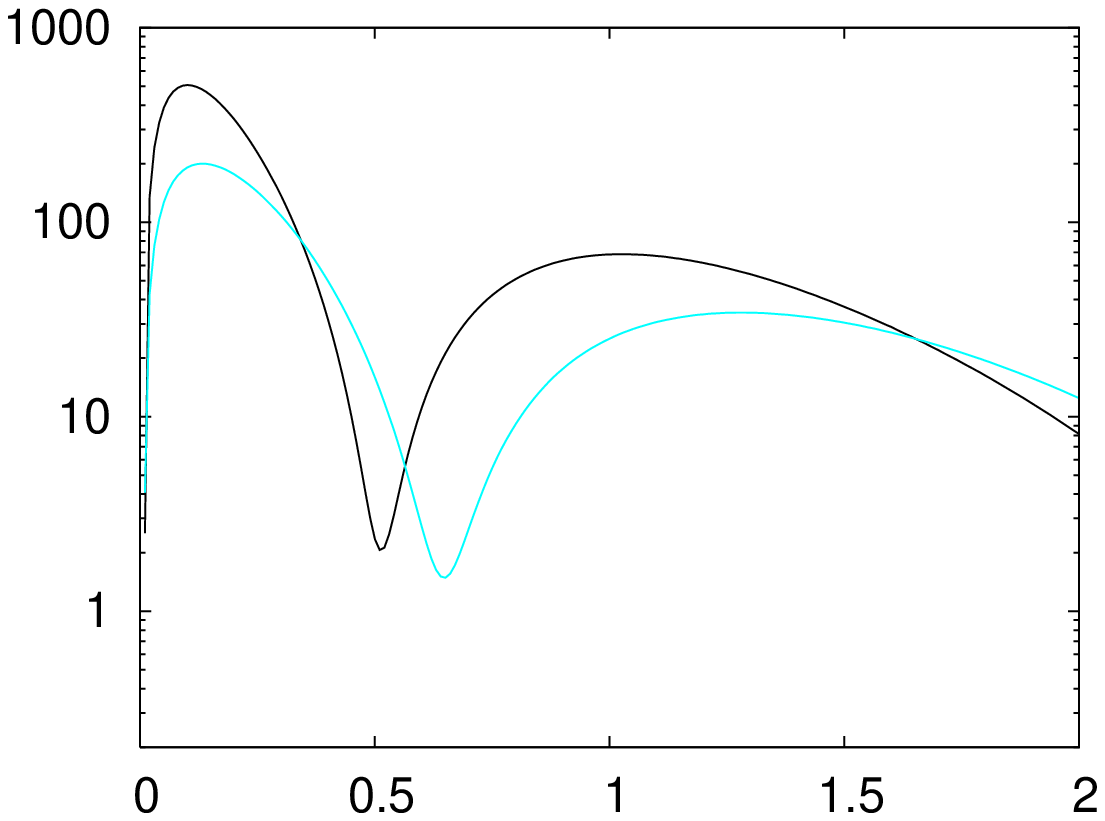}
\begin{picture}(0,0)
\setlength{\unitlength}{1mm}
\put(0,38){$\frac{d\sigma}{dt}$}
\put(50,45){\small{(a)}}
\put(40,1){\small{$-t$ (GeV$^2$)}}
\end{picture}
\end{minipage}
\begin{minipage}{70mm}
\epsfxsize70mm
\epsffile{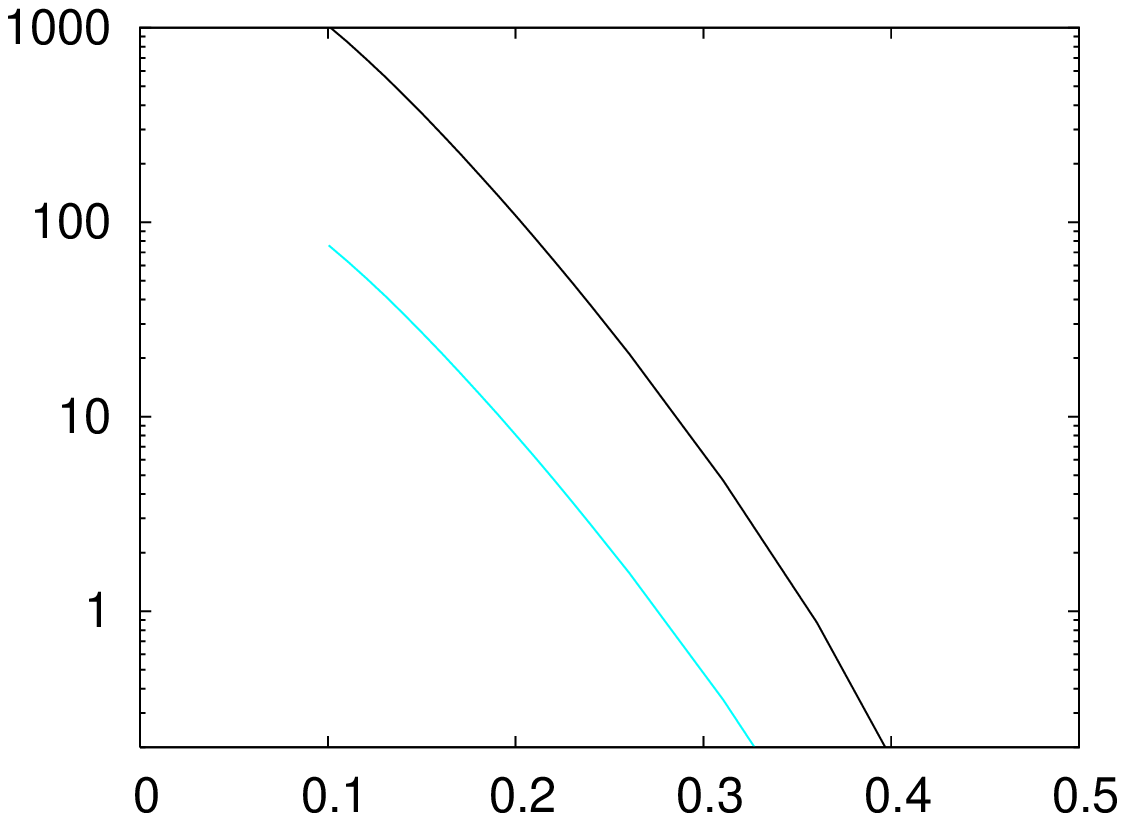}
\begin{picture}(0,0)
\setlength{\unitlength}{1mm}
\put(50,45){\small{(b)}}
\put(40,1){{\small $-t$ (GeV$^2$)}}
\end{picture}
\end{minipage}
\vskip 5truemm
\begin{minipage}{70mm}
\epsfxsize70mm
\epsffile{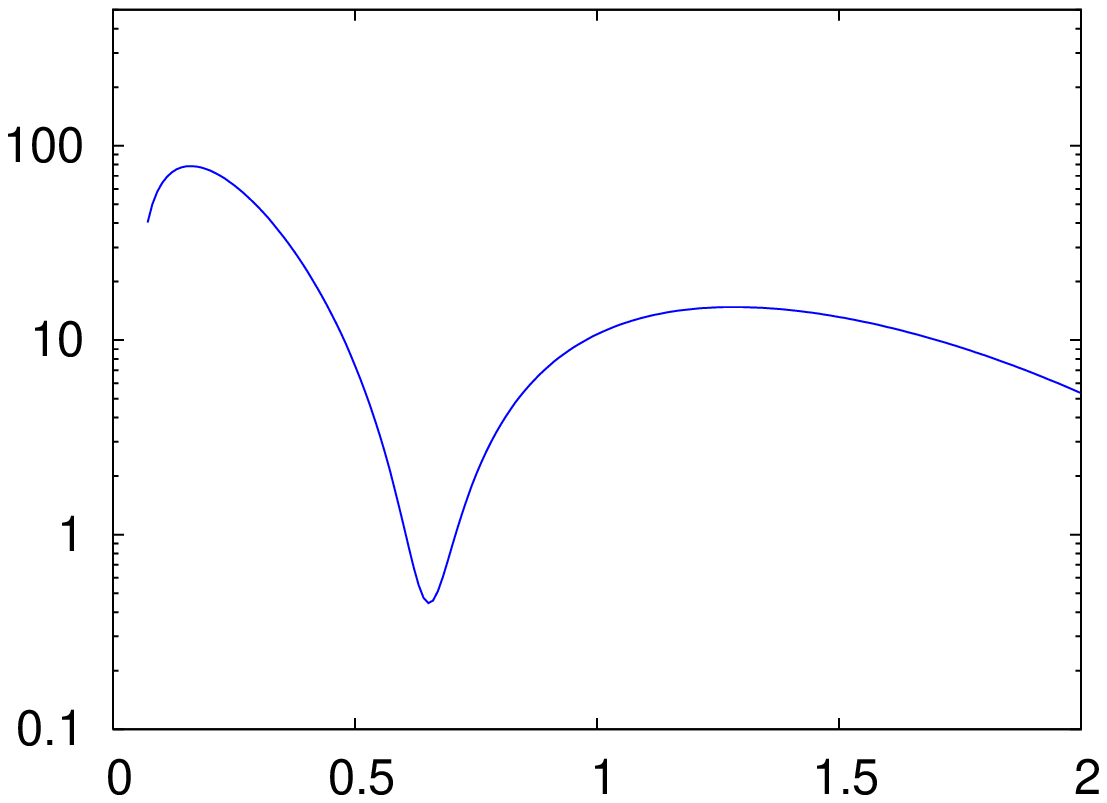}
\begin{picture}(0,0)
\setlength{\unitlength}{1mm}
\put(0,38){$\frac{d\sigma}{dt}$}
\put(50,45){\small{(c)}}
\put(40,1){{\small $-t$ (GeV$^2$)}}
\end{picture}
\end{minipage}
\begin{minipage}{70mm}
\epsfxsize70mm
\epsffile{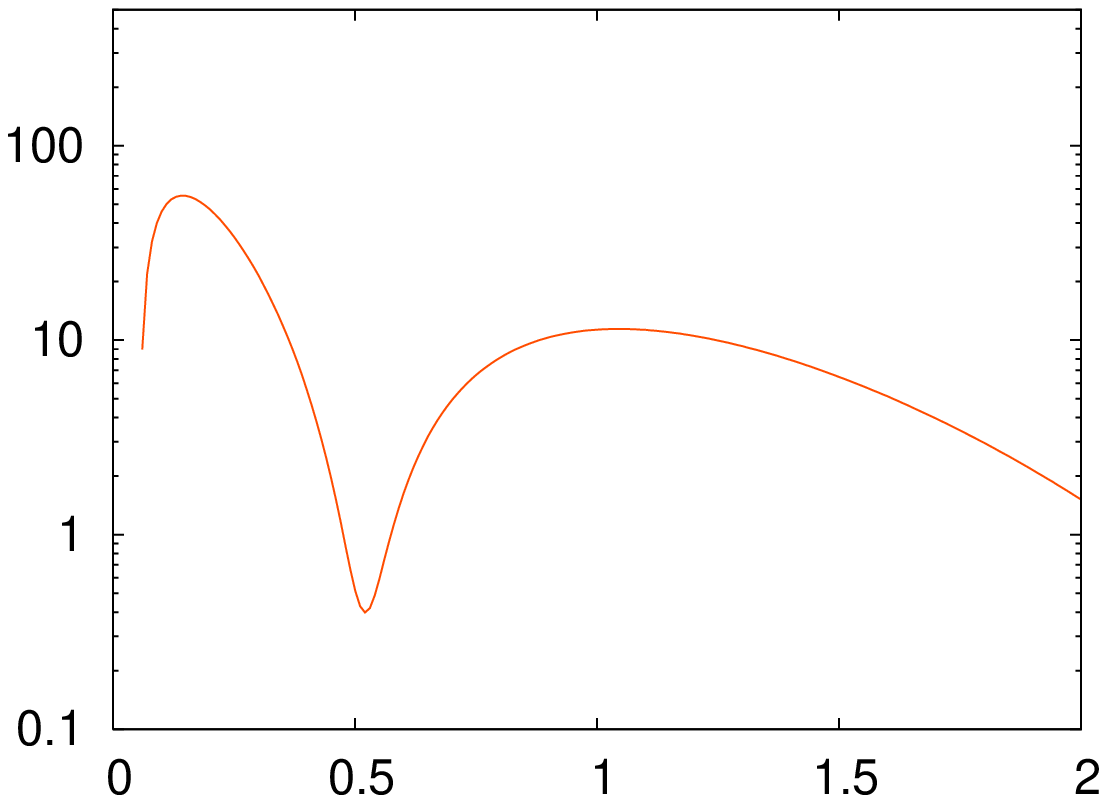}
\begin{picture}(0,0)
\setlength{\unitlength}{1mm}
\put(50,45){\small{(d)}}
\put(40,1){{\small $-t$ (GeV$^2$)}}
\end{picture}
\end{minipage}
\caption{Differential photoproduction cross sections in nb GeV$^{-2}$
on (a) protons and (b) $^4$He for the $a_0(980)$ (brown curves) and
$f_0(980)$ (azure curves) assuming that they are members of the 
ground-state nonet and that the $f_0(980)$ is a pure singlet, and
on protons (c) for the $f_0(1500)$ assuming that it is a member of the
ground-state nonet and is a pure octet and (d) for the $a_0(1450)$ 
assuming it is in the first radial excitation.}
\label{dsigdt3}
\efig
\ec
\subsubsection{Scenario IV}
\label{scen4}

This scenario is analogous to scenario II with the $f_0(1370)$ replacing
the $f_0(1500)$ as the octet member of the ground-state nonet. The
integrated cross sections for photoproduction on protons and $^4$He at
$E_\gamma = 5$ GeV are $47$ nb and $1.1$ nb respectively. The 
differential cross section at $E_\gamma = 5$ GeV is shown in Figure 4(b).

\bfig[t]
\bc
\begin{minipage}{70mm}
\epsfxsize70mm
\epsffile{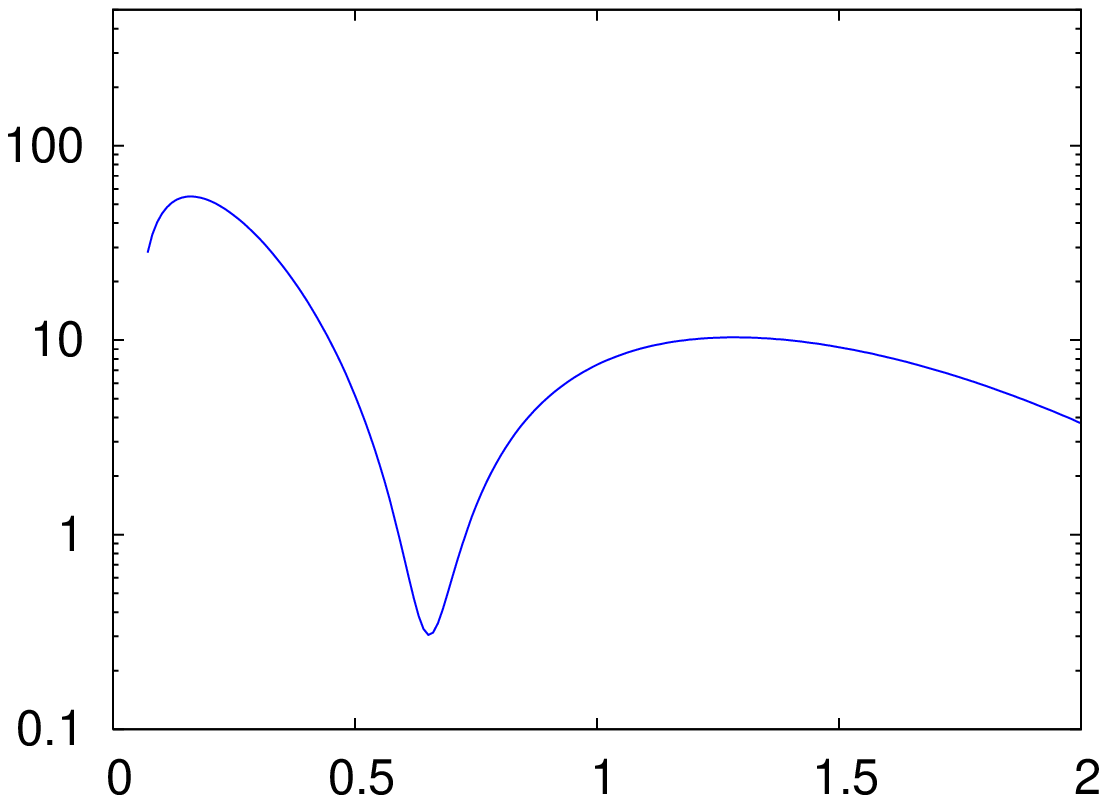}
\begin{picture}(0,0)
\setlength{\unitlength}{1mm}
\put(0,38){$\frac{d\sigma}{dt}$}
\put(55,45){{\small (a)}}

\put(40,1){{\small $-t$ (GeV$^2$)}}
\end{picture}
\end{minipage}
\begin{minipage}{70mm}
\epsfxsize70mm
\epsffile{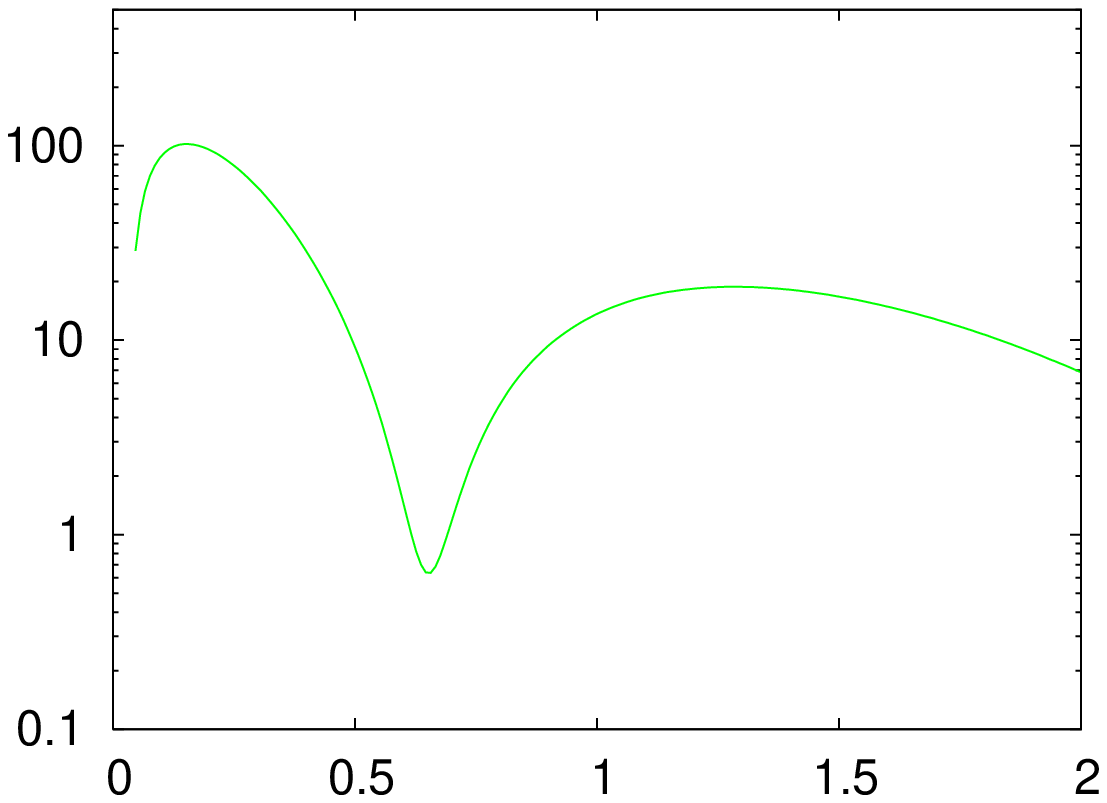}
\begin{picture}(0,0)
\setlength{\unitlength}{1mm}
\put(55,45){{\small (b)}}
\put(40,1){{\small $-t$ (GeV$^2$)}}
\end{picture}
\end{minipage}
\caption{Differential photoproduction cross sections on protons in nb
GeV$^{-2}$ for (a) $f_0(1500)$ assuming that it is a member of the
first radially-excited nonet and (b) $f_0(1370)$ assuming it is the
octet member of the ground state nonet.}
\ec
\label{dsigdt4}
\efig

\section{Mass Distributions}
\label{mass}

We present results for the $\pi^0\pi^0$ channel for the isoscalars. 
Other pseudoscalar-pseudoscalar channels can be obtained from 
these by scaling using the branching fractions of Table 7, or the results
of Bugg \cite{Bugg07} for the special case of the $f_0(1370)$.

\subsection{The signal}

To obtain mass distributions for the $f_0(1370)$, $f_0(1500)$, $f_0(1710)$ and
$a_0(1450)$, we represent them as relativistic Breit-Wigner resonances 
with energy-dependent partial widths. The signal cross section for the final 
state $i$ is given by (see Appendix)
\be
\frac{d\sigma}{dt~dM}=\frac{d\sigma_0(t,M)}{dt}\frac{2m_S^2}{\pi}
\frac{\Gamma_i(M)}{(m_S^2-M^2)^2+M^2\Gamma_{\rm Tot}^2}
\label{signaldcs}
\ee
where $d\sigma_0(t,M)/dt$ is the narrow-width differential cross 
section at a scalar mass $M$. For pseudoscalar-pseudoscalar final states the
partial width $\Gamma_i(M)$ is given in terms of the $SPP$ coupling $g_i$
by
\be 
\Gamma_i(M) = \frac{g_i^2 \rho(M,m_a,m_b)}{16\pi M},
\label{BW1}
\ee
with
\be 
\rho(M,m_a,m_b)=\sqrt{(1-(m_a+m_b)^2/M^2)(1-(m_a-m_b)^2/M^2)},
\label{kdef}
\ee
where $m_a$ and $m_b$ are the masses of the two scalars. Conversely $g_i$ 
is determined from the partial width at resonance, putting $M=m_S$ in 
(\ref{BW1}). 

The $4\pi$ channels $2\pi^+2\pi^-$, $\pi^+\pi^-2\pi^0$ and $4\pi^0$ represent 
a significant fraction of scalar decays and are dominated by $\rho\rho$ and 
$\sigma\sigma$. As we do not consider them explicitly, we represent them 
collectively using the parametrisation of $4\pi$ phase space suggested by 
Bugg \cite{Bugg07}:
\be
\rho_{4\pi}(M) =\frac{\sqrt{(1-16m_\pi^2/M^2)}}{1+\exp[-\Lambda(M^2-M_0^2)]}
\label{BW2}
\ee
with $M_0 = 1.799$ GeV and $\Lambda = 3.39$ GeV$^{-2}$. Then for the 
$4\pi$ states we take
\be
\Gamma_{4\pi} = \gamma_{4\pi}\frac{\rho_{4\pi}(M)}{\rho_{4\pi}(m_S)},
\label{BW3}
\ee
so that $\gamma_{4\pi}$ is the 4$\pi$ partial width at resonance.

The total width is then
\be
\Gamma_{\rm Tot}(M) = \sum_i \Gamma_i(M)
\label{BW4}
\ee 

Although the dominant decay of the $f_0(980)$ is $\pi\pi$ and that of the 
$a_0(980)$ is $\pi\eta$, both branching fractions being about 0.85 with the 
remainder in $K\bar{K}$ \cite{PDG06}, these states reside at the 
$K \bar K$ threshold, so the procedure outlined above is not reliable. 
Instead we use the Breit-Wigner parametrisations obtained in the analysis 
of $\phi$ radiative decays \cite{KLOE07f,KLOE07}. In this section the 
"no-structure" versions of the fits are employed, which correspond to a
point-like $\phi \gamma S$ vertex, and are in line with the quark-loop 
radiative transition assumption. The Breit-Wigner width takes the form
\beqa
\Gamma(M) &=& g_{\pi\pi}^2\frac{v_\pi(M)}{8\pi M^2}+\nonumber\\
&& g_{K\bar{K}}^2\frac{v_{K^\pm}(M)+v_{K^0}(M)}{8\pi M^2},
\label{kloe}
\eeqa
where $v_{\pi}(M)=\sqrt{M^2/4-M^2_{\pi\pi}}$ and $v_{K}(M)=\sqrt{M^2/4-
M^2_{K\bar{K}}}$ are momenta with an analytical continuation
below threshold. The corresponding parameters are
$M=0.9847$ GeV, $g_{K^+K^-}=g_{K^0 \bar 
K^0}=0.4$ GeV, $g_{\pi^+\pi^-}=\sqrt{2}g_{\pi^0\pi^0}=1.31$ GeV for the 
$f_0(980)$, and $M=0.983$ GeV, $g_{K^+K^-}=g_{K^0 \bar K^0}=1.57$ GeV, 
$g_{\pi\eta}=2.2$ GeV for the $a_0(980)$.

\begin{table}
\bc
\btab{|l|c|c|c|}
\hl
Channel & PDG & WA102/CK & CB \\
\hl
$\pi\pi$          & $34.9 \pm 2.3$ & $33.7 \pm  3.4$ & $33.9 \pm  3.7$ \\
$\eta\eta$        & $ 5.1 \pm 0.9$ & $ 6.1 \pm  0.1$ & $ 2.6 \pm  0.3$ \\
$\eta\eta^\prime$ & $ 1.9 \pm 0.8$ & $ 3.2 \pm  0.7$ & $ 2.2 \pm  0.1$ \\
$K\bar{K}$        & $ 8.6 \pm 0.1$ & $10.7 \pm  2.4$ & $ 6.2 \pm  0.5$ \\
$4\pi$            & $49.5 \pm 3.3$ & $46.3 \pm  8.5$ & $55.1 \pm 16.9$ \\
\hl
\etab
\caption{Branching fractions in percent for the $f_0(1500)$ from the
PDG \cite{PDG06}, the WA102 experiment \cite{WA102} from the analysis of 
Close
and Kirk \cite{CK01} (CK) and the Crystal Barrel experiment \cite{CB01} 
(CB).}
\ec
\label{1500bf}
\end{table}

In order to obtain mass distributions it is necessary to have accurate
branching fractions to hadronic final states. This is the case for the
$f_0(1500)$ but not for the $f_0(1370)$ or $f_0(1710)$, particularly 
the former \cite{PDG06,Ochs06,Bugg07,KZ07,AO08}.
The various hadronic decay channels of the 
$f_0(1500)$ are well defined. This is illustrated in Table 6
in which the
branching fractions, in percent, are given from the PDG \cite{PDG06},
the WA102 experiment \cite{WA102} as obtained in the analysis of Close
and Kirk \cite{CK01} and the Crystal Barrel experiment \cite{CB01}.
The usefulness of the $a_0(1450)$ as a check on the model is also compromised
by the limited information on the hadronic branching fractions 
\cite{PDG06} which we discuss below.

For definiteness we use the results of the WA102 collaboration \cite{WA102}
for the isoscalars. These comprise a complete data set for the decay of 
the $f_0(1370)$, $f_0(1500)$ and $f_0(1710)$ to all pseudoscalar meson 
pairs. As only relative branching ratios are provided, to get the 
absolute branching ratios that we require we use the analysis of these 
data in \cite{CK01} to take account of other channels. The branching 
fractions to pseudoscalars are summarised in Table 7.
Of course for the $\pi^0\pi^0$ channel
the $\pi\pi$ branching fraction shown has to be divided by a factor of 
three. We also take results of the WA102 collaboration \cite{WA102} for the 
total widths, $272 \pm 50$ MeV for the $f_0(1370)$, $108 \pm 18$ MeV for 
the $f_0(1500)$ and $124 \pm 24 $ MeV for the $f_0(1710)$.

We see from Table 7
that the branching fractions of the
$f_0(1370)$ to pseudoscalars are small, the principal decay mode being to
$4\pi$ \cite{CB01}. However there is a major disagreement with the 
$\pi\pi$ branching fraction shown in Table 7.
In the analysis of Bugg 
\cite{Bugg07}, the $2\pi$:$4\pi$ ratio at resonance is given as 6:1. The 
results of Albaladejo and Oller \cite{AO08} imply that the three 
pseudoscalar channels $\pi\pi$, $\eta\eta$ and $K\bar{K}$ saturate the 
decay modes of the $f_0(1370)$ with $\pi\pi$ dominant. The consequences 
of this alternative view of the $f_0(1370)$ are discussed where 
appropriate.

The relative branching fractions $\pi\eta^\prime(980)/\pi\eta$ and
$K\bar{K}/\pi\eta$ of the $a_0(1450)$ are $0.35 \pm 0.16$ and $0.88 \pm 
0.23$ respectively \cite{PDG06}. However the dominant decay mode of the 
$a_0(1450)$ appears to be $\omega\pi\pi$ \cite{CB03}, although there is 
some uncertainty in the actual branching fraction \cite{PDG06}. Relative 
to $\pi\eta$ it is quoted as  $10.7 \pm 2.3$, obtained by comparing the 
total rates of $p\bar{p} \to a_0(1450)\pi$ for $a_0(1450) \to \pi\eta$ 
and $a_0(1450) \to \omega\pi^+\pi^-$ and assuming the $\omega\pi\pi$ 
final state is $\omega\rho$.
The uncertainty is also reflected in the width, so as the $a_0(1450)$ is
peripheral to our argument we do not show any mass distributions but 
simply note that the cross section for $a_0(1450)$ photoproduction is 
sufficiently large for it to be used to clarify the $a_0(1450)$ decay modes.

\begin{table}
\bc
\btab{|c|c|c|c|c|}
\hl
State & $\pi\pi$ & $K\bar{K}$ & $\eta\eta$ & $\eta\eta^\prime$  \\
\hl
$f_0(1370)$ & 0.027 & 0.013 & 0.004 & \\
$f_0(1500)$ & 0.337 & 0.107 & 0.061 & 0.032 \\
$f_0(1710)$ & 0.119 & 0.595 & 0.286 & \\
\hl
\etab
\caption{Branching fractions for the scalars from \cite{WA102} and
\cite{CK01}.}
\ec
\label{bratios1}
\end{table}

\bfig
\bc
\begin{minipage}{70mm}
\epsfxsize70mm
\epsffile{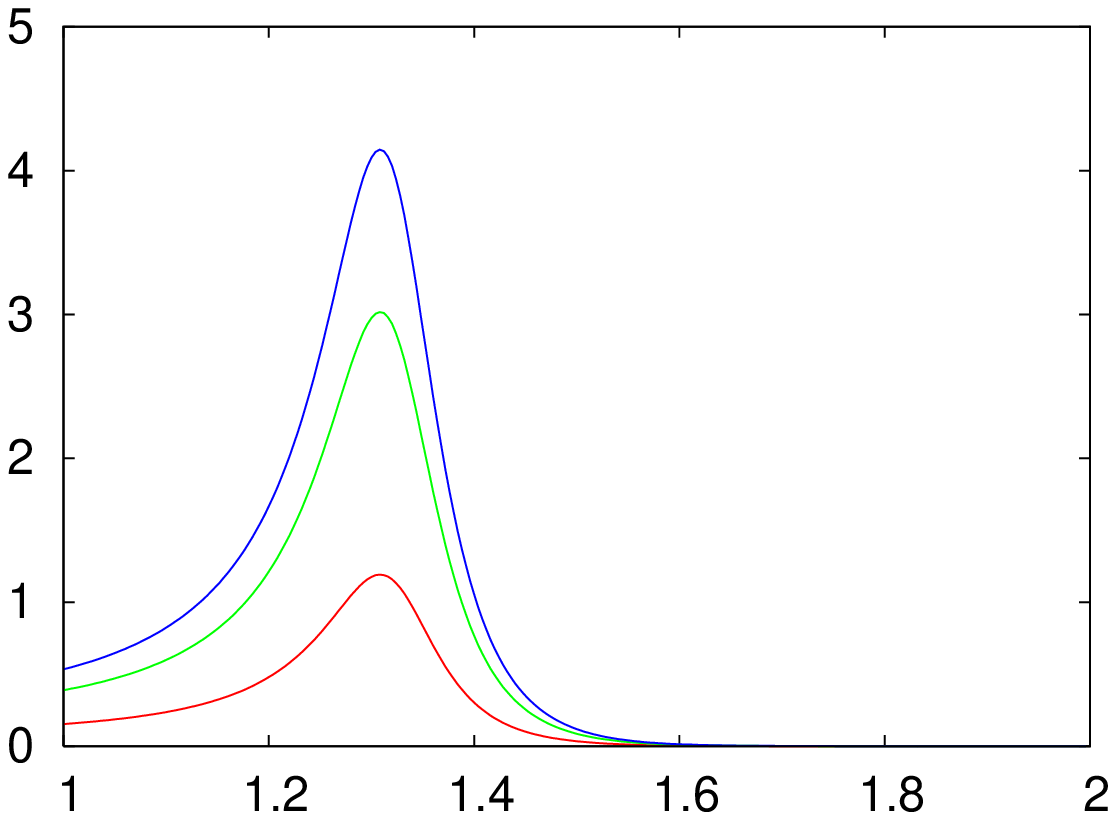}
\begin{picture}(0,0)
\setlength{\unitlength}{1mm}
\put(0,38){$\frac{d\sigma}{dM}$}
\put(55,45){{\small (a)}}
\put(40,1){{\small $M$ (GeV)}}
\end{picture}
\end{minipage}
\hfill
\begin{minipage}{70mm}
\epsfxsize70mm
\epsffile{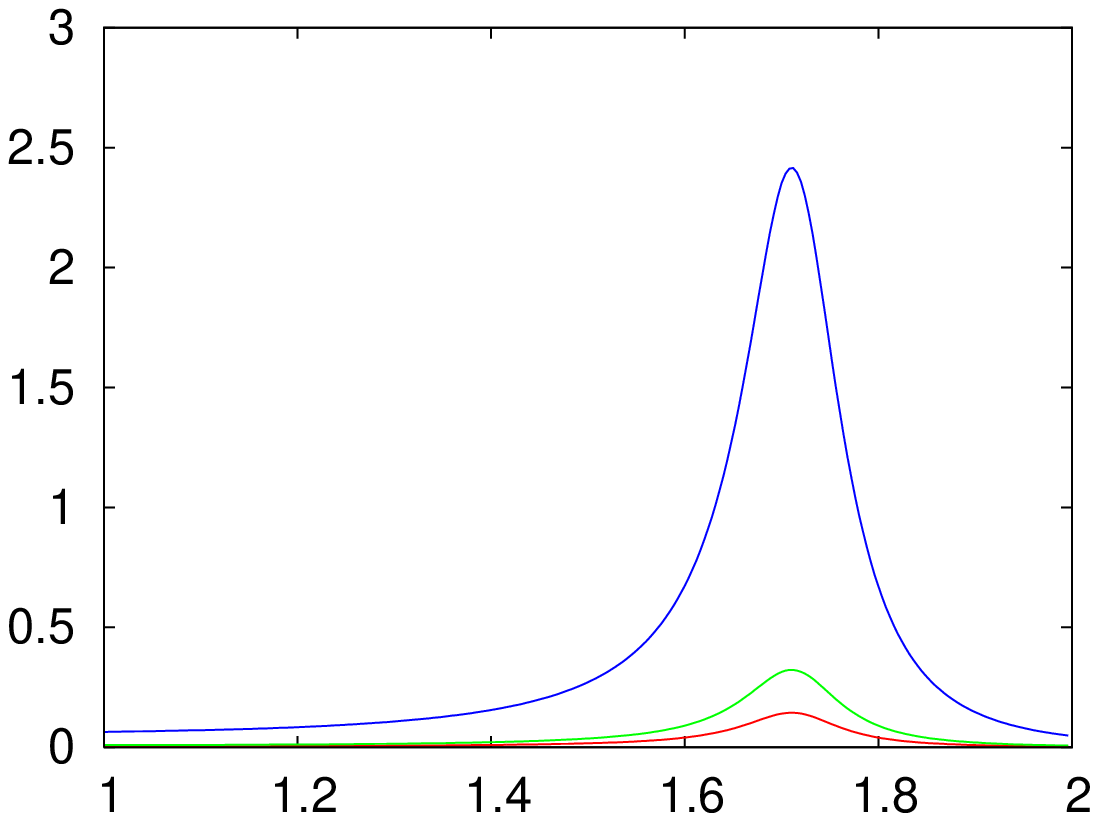}
\begin{picture}(0,0)
\setlength{\unitlength}{1mm}
\put(55,45){{\small (b)}}
\put(40,1){{\small $M$ (GeV)}}
\end{picture}
\end{minipage}
\vskip5truemm
\begin{minipage}{70mm}
\epsfxsize70mm
\epsffile{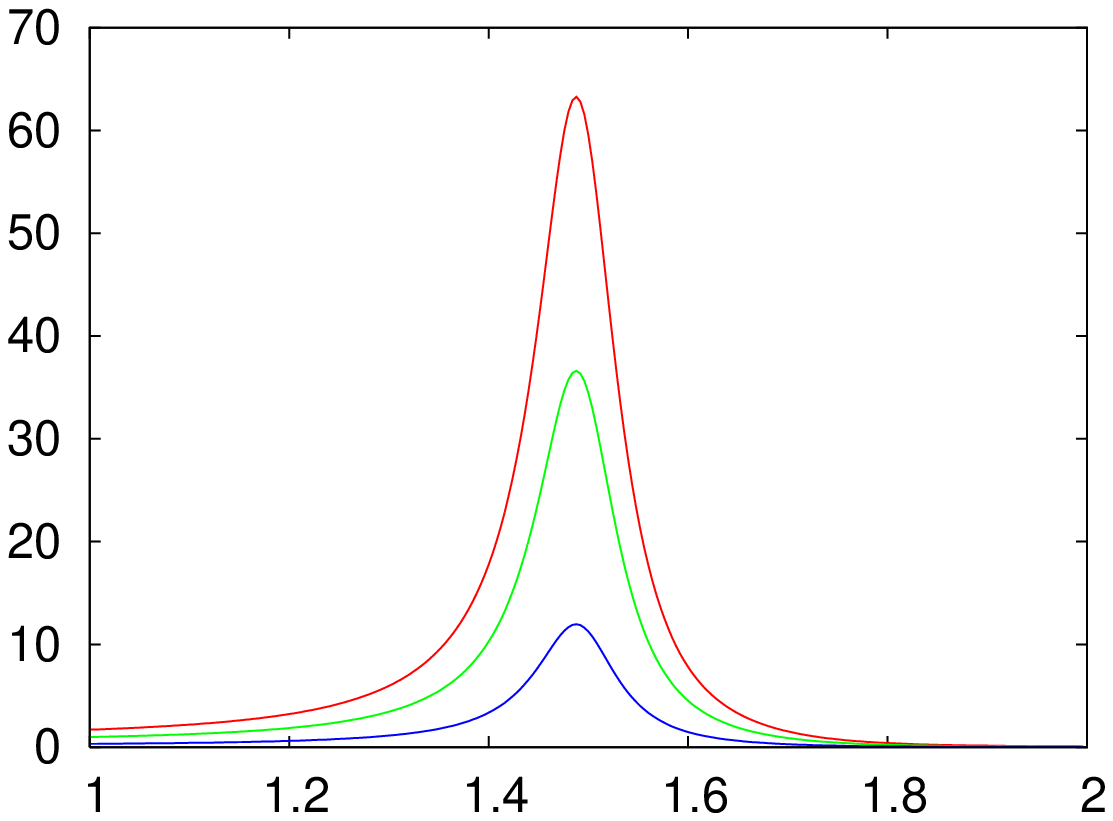}
\begin{picture}(0,0)
\setlength{\unitlength}{1mm}
\put(55,45){{\small (c)}}
\put(-2,40){$\frac{d\sigma}{dM}$}
\put(40,1){{\small $M$ (GeV)}}
\end{picture}
\end{minipage}
\caption{Scenario I. Differential $\pi^0\pi^0$ mass distributions in nb
GeV$^{-1}$ at $E_\gamma = 5$ GeV for (a) $f_0(1370)$, (b) $f_0(1710)$ and
(c) $f_0(1500)$. The glueball masses are L (red), M (green) and H (blue) 
in
each figure.}
\label{dmass1}
\ec
\efig

\bfig
\bc
\begin{minipage}{60mm}
\epsfxsize60mm
\epsffile{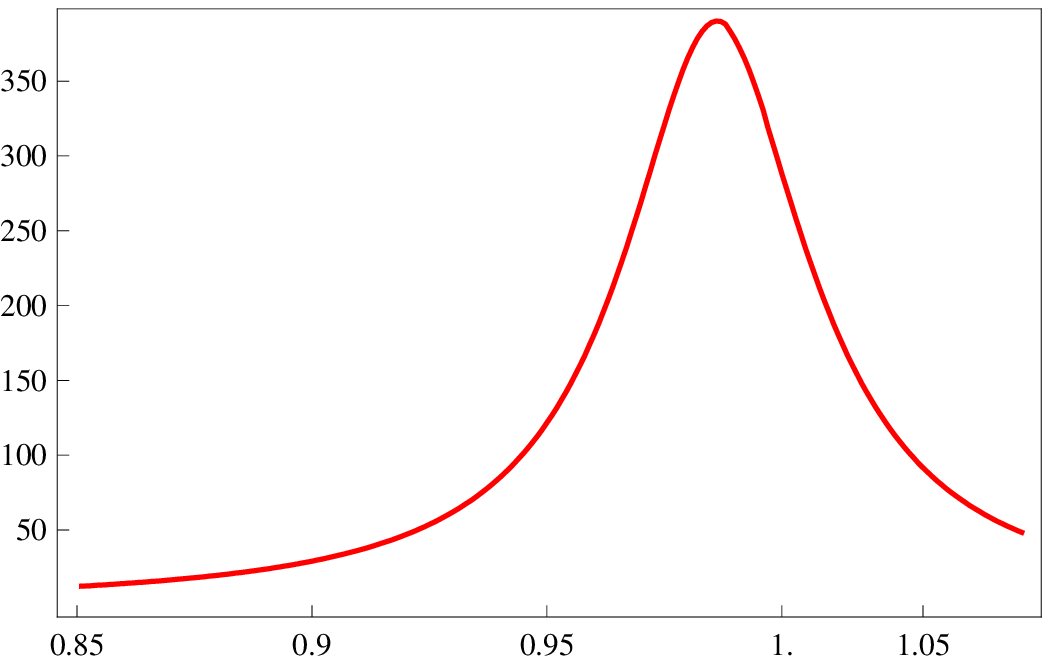}
\begin{picture}(0,0)
\setlength{\unitlength}{1mm}
\put(-5,35){$\frac{d\sigma}{dM}$}
\put(10,35){{\small (a)}}
\put(40,1){\small{$M$ (GeV)}}
\end{picture}
\end{minipage}
\hskip5truemm
\begin{minipage}{60mm}
\epsfxsize60mm
\epsffile{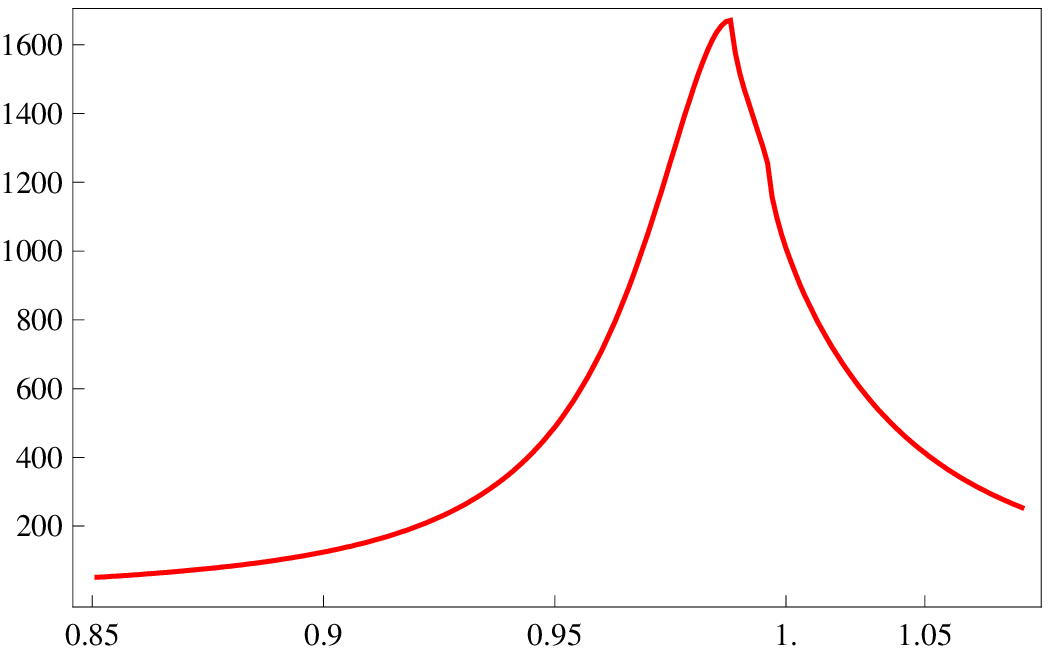}
\begin{picture}(0,0)
\setlength{\unitlength}{1mm}
\put(40,1){{\small $M$ (GeV)}}
\put(10,35){{\small (b)}}
\end{picture}
\end{minipage}
\caption{Scenario II. (a) Differential $\pi^0\pi^0$ mass distribution in nb 
GeV$^{-1}$ for the $f_0(980)$ at $E_\gamma = 5$ GeV.
(b)  Differential $\pi^0\eta^0$ mass distributions in nb GeV$^{-1}$ for the
neutral $a_0(980)$ at $E_\gamma = 5$ GeV.}
\label{dmass2}
\ec
\efig

\bc
\bfig
\begin{minipage}{70mm}
\epsfxsize70mm
\epsffile{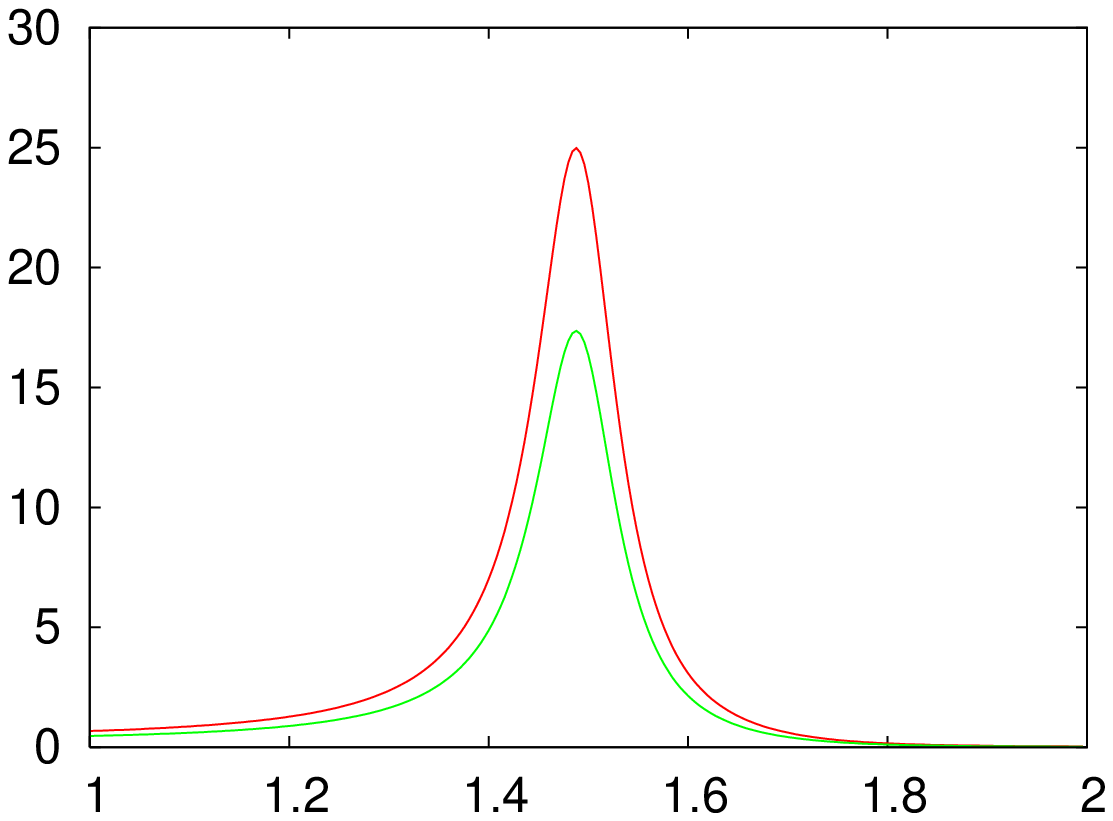}
\begin{picture}(0,0)
\setlength{\unitlength}{1mm}
\put(-2,36){$\frac{d\sigma}{dM}$}
\put(55,45){{\small (a)}}
\put(40,1){{\small $M$ (GeV)}}
\end{picture}
\end{minipage}
\begin{minipage}{70mm}
\epsfxsize70mm
\epsffile{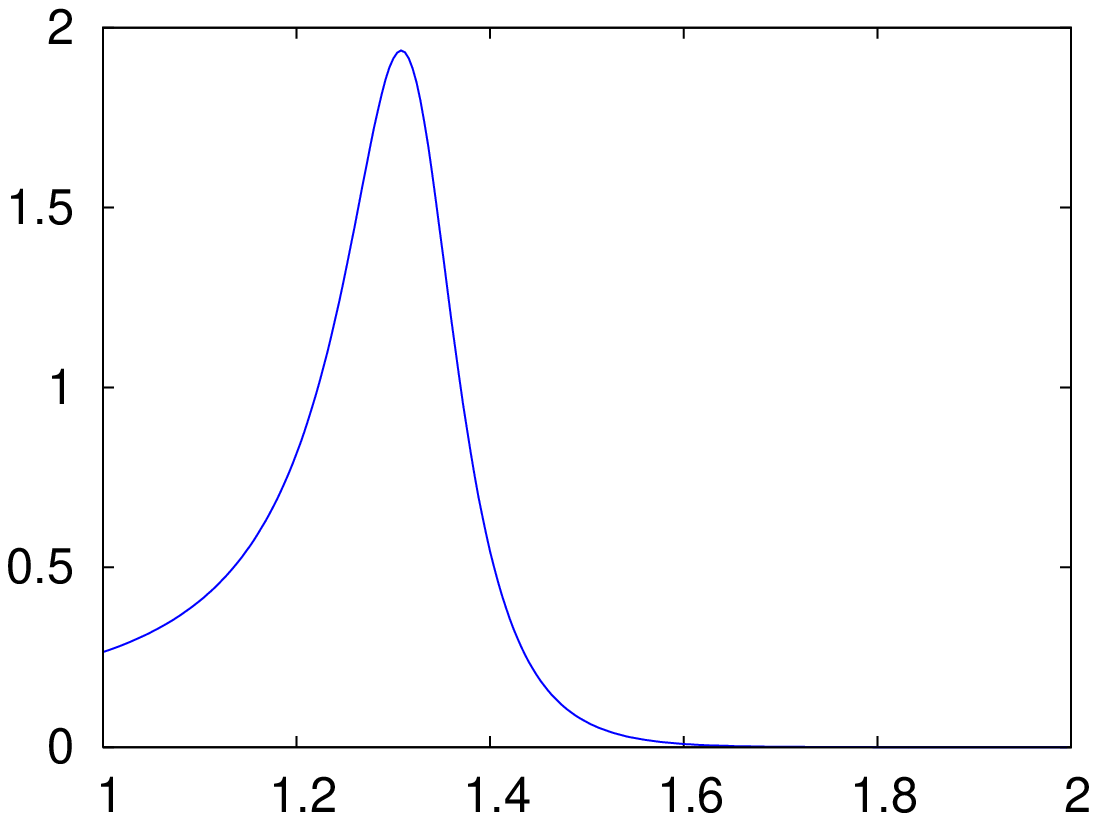}
\begin{picture}(0,0)
\setlength{\unitlength}{1mm}
\put(55,45){{\small (b)}}
\put(40,1){{\small $M$ (GeV)}}
\end{picture}
\end{minipage}
\caption{Scenarios III and IV. (a) Differential $\pi^0\pi^0$ mass 
distributions
in nb GeV$^{-1}$ at $E_\gamma = 5$ GeV for the $f_0(1500)$ as an octet
ground state (red) and as a member of the first radial excitation 
(green).
(b) Differential $\pi^0\pi^0$ mass distribution in nb GeV$^{-1}$ at
$E_\gamma = 5$ GeV for the $f_0(1370)$ as an octet ground state.}
\label{dmass3}
\efig
\ec

\subsubsection{Scenario I}

The $\pi^0\pi^0$ mass distributions $d\sigma/dM$ for the $f_0(1370)$,
$f_0(1500$ and the $f_0(1710)$ are given in Figure 5,
in each case for light, medium and heavy glueball masses and using the
branching fractions of Table 7. Note the difference in scale in Figure 
5(c), reflecting the small $\pi\pi$ branching fraction for the $f_0(1370)$ 
and the small cross section for the $f_0(1710)$. However if the $\pi\pi$ 
branching fraction of the $f_0(1370)$ from the analysis of \cite{Bugg07,AO08}
is used instead of that in Table 7 the mass distribution is about a 
factor of 30 larger. The distortion of the Breit-Wigner line shape for 
the $f_0(1370)$ arises primarily from the combined effect of the small 
$\pi\pi$ branching fraction and the large, rapidly-rising $4\pi$ channel in 
the denominator. As the $f_0(980)$ and $a_0(980)$ are not $n\bar{n}$ states
in this scenario discussion of them is deferred to Section 5, where the
extreme possibility of a purely molecular assignment is considered.

\subsubsection{Scenario II}

The $a_0(980)$, $f_0(980)$ and $f_0(1500)$ are members of the 
ground-state nonet with the $f_0(980)$ and $f_0(1500)$ mixed such that 
the former is a singlet and the latter is an octet. The $\pi^0\pi^0$ and 
$\pi^0\eta^0$ mass distributions for the $f_0(980)$ and $a_0(980)$ are 
shown in Figure 6(a) and 6(b) respectively. The $\pi^0\pi^0$ mass 
distribution for the $f_0(1500)$ is given as the red curve in Figure 
7(a). The $f_0(1370)$ does not exist in this scenario.

\subsubsection{Scenario III}

The $a_0(980)$ and the $f_0(980)$ are as in scenario II, although the 
latter is no longer a singlet so the $\pi^0\pi^0$  mass distribution 
should be scaled up by a factor of 1.5. The $f_0(1500)$ is now a member 
of the first radial excitation and the corresponding $\pi^0\pi^0$ mass 
distributionm is shown as the green curve in Figure 7(a). The $f_0(1370)$ 
does not exist in this scenario.

\subsubsection{Scenario IV}

This is analogous to scenario II, but the octet member is the $f_0(1370)$.
The $\pi^0\pi^0$ mass distribution is given in Figure 7(b) using the
branching fraction of Table 7. Recall that if the $\pi\pi$ branching
fraction of the $f_0(1370)$ from the analysis of \cite{Bugg07,AO08} is 
used instead of that in Table 7 the mass distribution is about a factor 
of 30 larger. The $f_0(1500)$ and $f_0(1710)$ are considered to be unmixed
glueballs so cannot be photoproduced directly.

\subsection{Continuum background}
\label{continuum}

There is a coherent, continuum background in the $\pi^0\pi^0$, 
$\pi^0\eta^0$ and $\eta^0\eta^0$ channels. The production mechanism is 
illustrated in Figure 8.  

Each graph has the form
\be
M_\mu = g_1g_2\epsilon_{\mu\rho\beta\gamma}q_{\beta}v_{\gamma}
\epsilon_{\rho\nu\lambda\sigma}p_{\lambda}v_{\sigma}F_{\nu}{\it
D}(s,t)\Pi(v).
\label{backamp1}
\ee
As before, $q$ is the photon momentum, $p = p_1-p_2$ where $p_1$ and
$p_2$ are the initial and final proton momenta and $v = k-q$ where
$k$ is the momentum of the pseudoscalar in the upper vertex. $F_\nu$ is
given by (\ref{gsv}), $D(s,t)$ is the Regge propagator (\ref{regge}) and
\be
\Pi(v) = \frac{1}{m_v^2-v^2-im_v\Gamma_v},
\label{bwprop}
\ee
where $m_v$ and $\Gamma_v$ are the mass and width of the vector meson in
the upper half of the graph. The quantities $g_1$, $g_2$ are respectively
the $\gamma$-$P_1$-$V_1$ and $V_1$-$P_2$-$V_2$ coupling constants at the top
and middle vertices of the diagrams.

\bfig
\bc
\begin{minipage}{30mm}
\epsfxsize30mm
\epsffile{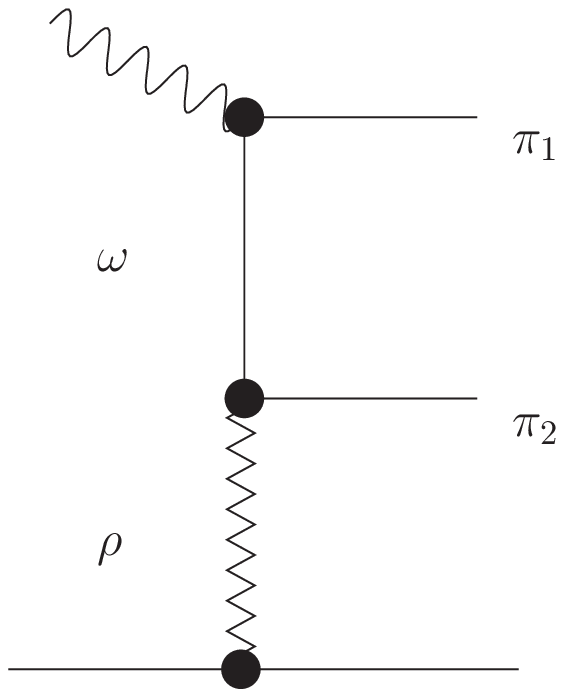}
\begin{picture}(0,0)
\setlength{\unitlength}{1mm}
\end{picture}
\end{minipage}
\begin{minipage}{30mm}
\epsfxsize30mm
\epsffile{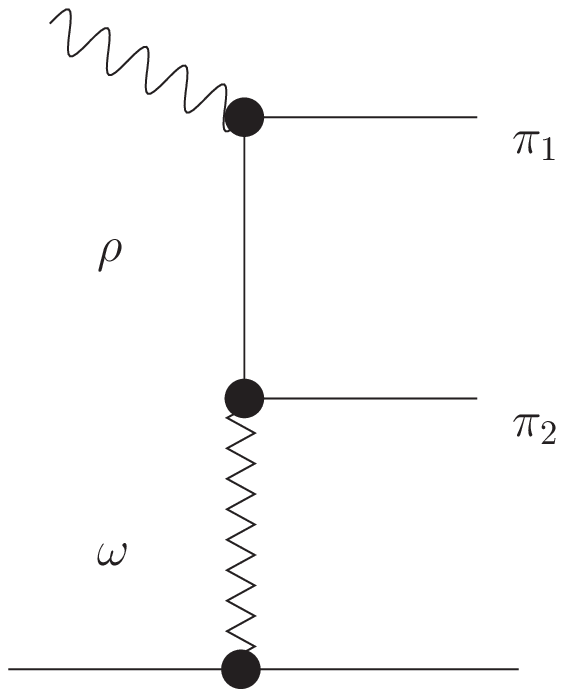}
\begin{picture}(0,0)
\setlength{\unitlength}{1mm}
\end{picture}
\end{minipage}
\hfill
\begin{minipage}{30mm}
\epsfxsize30mm
\epsffile{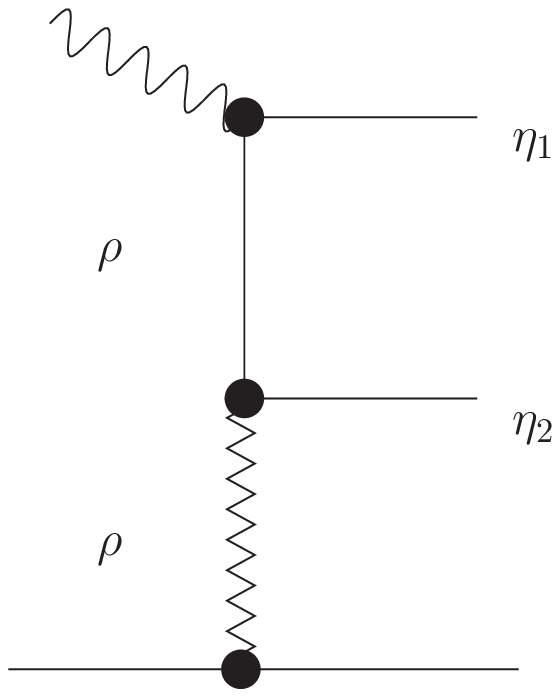}
\begin{picture}(0,0)
\setlength{\unitlength}{1mm}
\end{picture}
\end{minipage}
\begin{minipage}{30mm}
\epsfxsize30mm
\epsffile{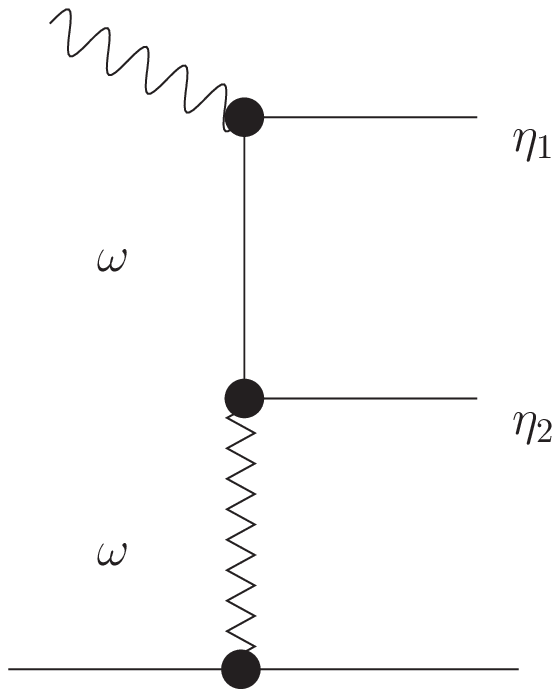}
\begin{picture}(0,0)
\setlength{\unitlength}{1mm}
\end{picture}
\end{minipage}
\vskip5truemm
\begin{minipage}{30mm}
\epsfxsize30mm
\epsffile{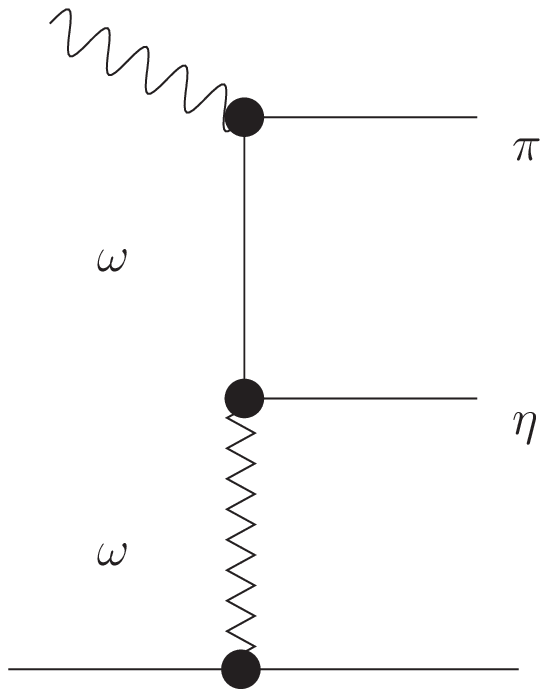}
\begin{picture}(0,0)
\setlength{\unitlength}{1mm}
\end{picture}
\end{minipage}
\begin{minipage}{30mm}
\epsfxsize30mm
\epsffile{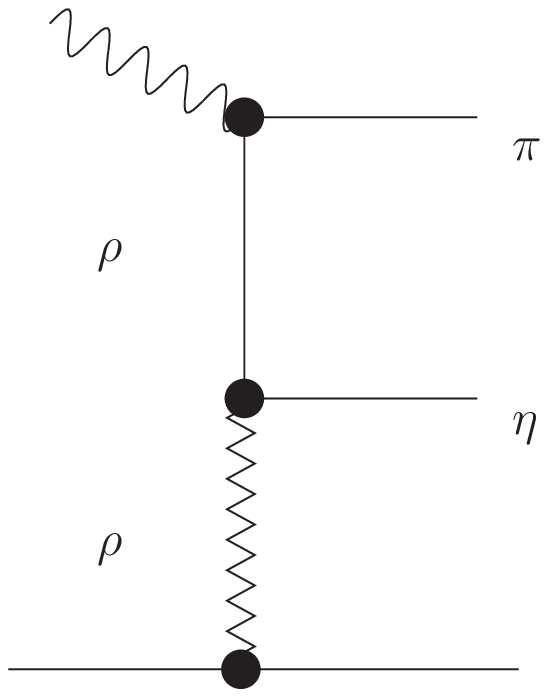}
\begin{picture}(0,0)
\setlength{\unitlength}{1mm}
\end{picture}
\end{minipage}
\begin{minipage}{30mm}
\epsfxsize30mm
\epsffile{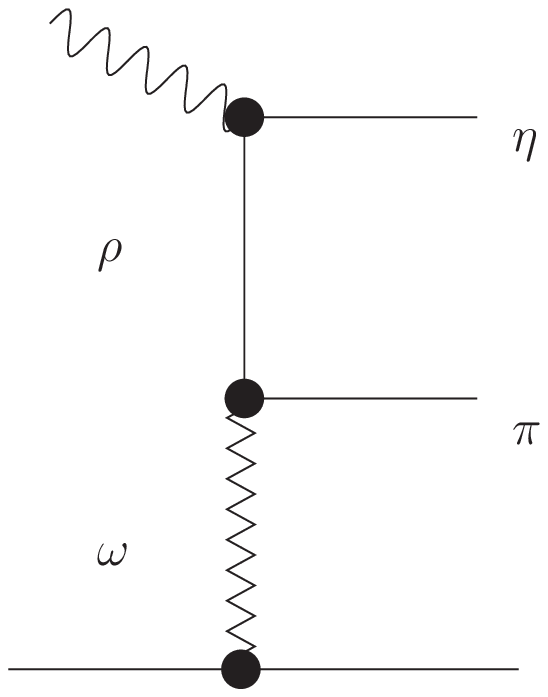}
\begin{picture}(0,0)
\setlength{\unitlength}{1mm}
\end{picture}
\end{minipage}
\begin{minipage}{30mm}
\epsfxsize30mm
\epsffile{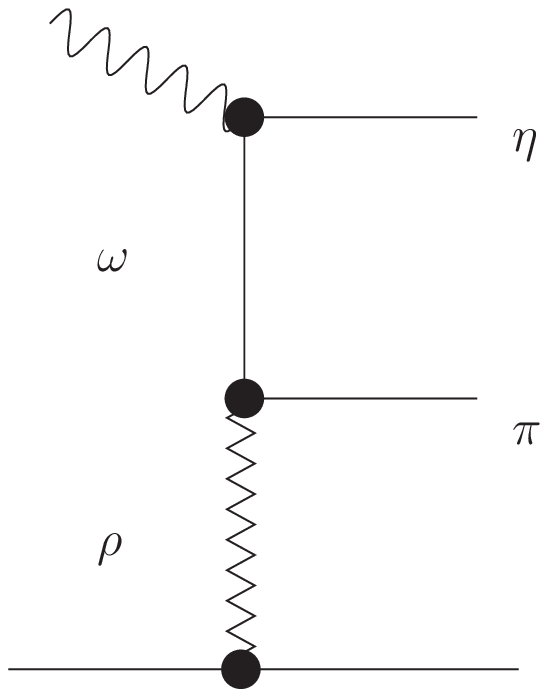}
\begin{picture}(0,0)
\setlength{\unitlength}{1mm}
\end{picture}
\end{minipage}
\ec
\label{Feynman}
\caption{Feynman diagrams for continuum $\pi^0\pi^0$, $\eta^0\eta^0$ and
$\pi^0\eta^0$ photoproduction. In addition there are diagrams with
$\pi_1 \leftrightarrow \pi_2$ and $\eta_1 \leftrightarrow \eta_2$ for
the first two.}
\efig

\bfig
\bc
\begin{minipage}{60mm}
\epsfxsize60mm
\epsffile{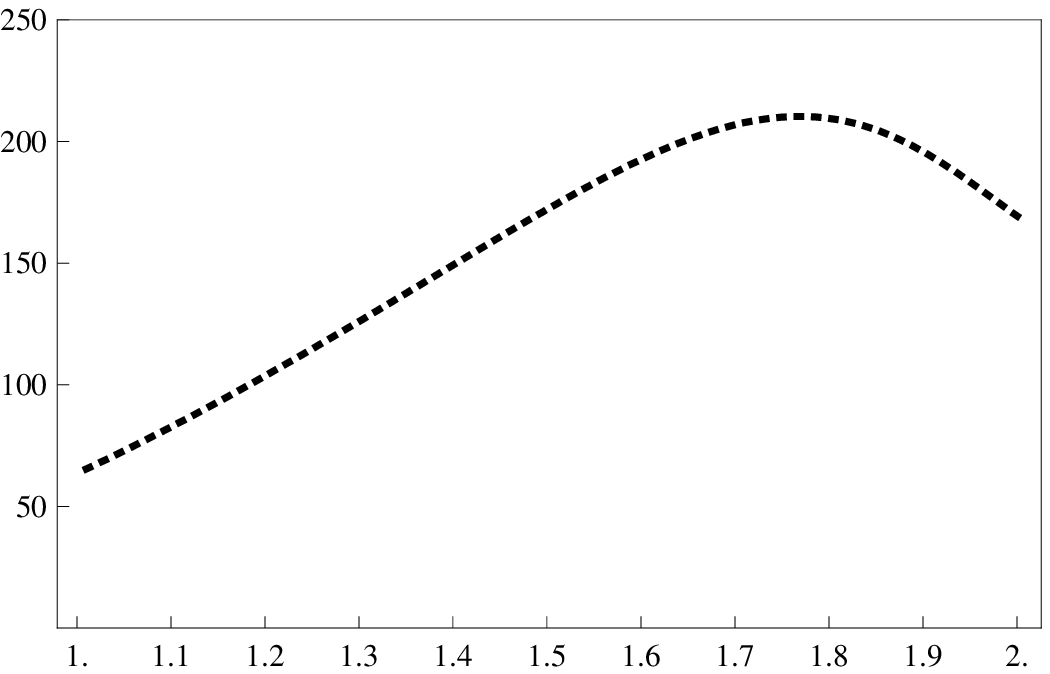}
\begin{picture}(0,0)
\setlength{\unitlength}{1mm}
\put(15,30){{\small $\pi^0\pi^0$}}
\put(-6,36){$\frac{d\sigma}{dM}$}
\put(40,1){{\small$M$ (GeV)}}
\end{picture}
\end{minipage}
\hskip1truecm
\begin{minipage}{60mm}
\epsfxsize60mm
\epsffile{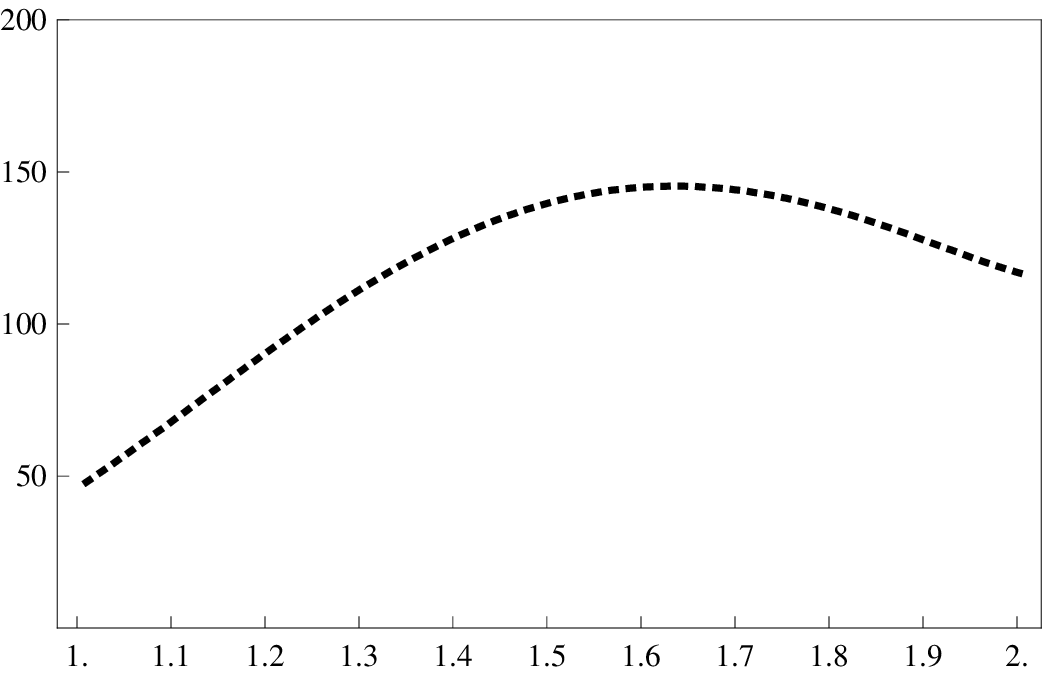}
\begin{picture}(0,0)
\setlength{\unitlength}{1mm}
\put(15,30){{\small $\pi^0\eta^0$}}
\put(-6,36){$\frac{d\sigma}{dM}$}
\put(40,1){{\small$M$ (GeV)}}
\end{picture}
\end{minipage}
\vskip5truemm
\begin{minipage}{60mm}
\epsfxsize60mm
\epsffile{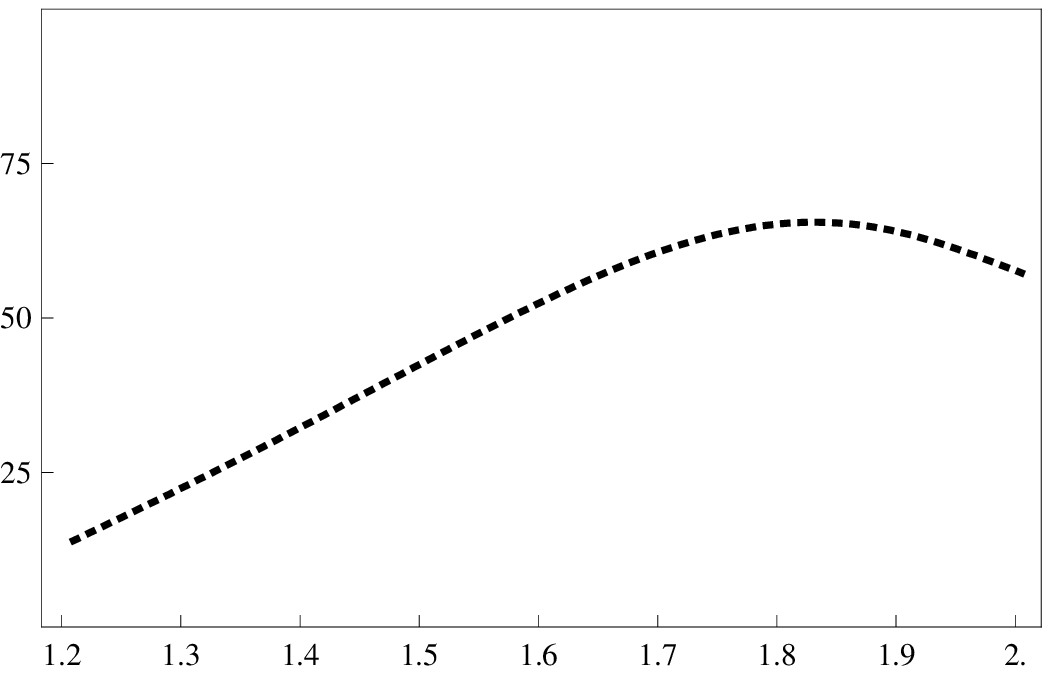}
\begin{picture}(0,0)
\setlength{\unitlength}{1mm}
\put(15,30){{\small $\eta^0\eta^0$}}
\put(-6,36){$\frac{d\sigma}{dM}$}
\put(40,1){{\small$M$ (GeV)}}
\end{picture}
\end{minipage}
\ec
\label{back}
\caption{Continuum $\pi^0\pi^0$, $\pi^0\eta^0$ and $\eta^0\eta^0$
backgrounds
in nb GeV$^{-1}$.}
\efig
 
The product of the coupling constants $g_1g_2 \equiv Cg_0$, where 
$g_0$ can be estimated from \cite{Bramon92}:
\be
g_0 =\frac{G^2e}{g\sqrt{2}},~~~~G=\frac{3\sqrt{2}g^2}{4\pi^2f},
\label{g01}
\ee
with $g = 4.2$ and $f= 132$ MeV. This gives $g_0^2 = 1.514\times 10^4
\alpha_{\rm em}$ GeV$^{-2}$. The constant $C=C_{V_2P_1P_2}$ is the 
appropriate  Clebsch-Gordan coefficient,
\be
1=C_{\rho\pi\pi}=3C_{\omega\pi\pi}=3{\textstyle\sqrt{\frac{3}{2}}}
C_{\rho\pi\eta}=
{\textstyle\sqrt{\frac{3}{2}}}C_{\omega\pi\eta}
={\textstyle\frac{3}{2}}C_{\rho\eta\eta}={\textstyle\frac{9}{2}}
C_{\omega\eta\eta}.
\ee
The explicit calculation is given in the Appendix and the results for
$\pi^0\pi^0$, $\pi^0\eta^0$ and $\eta^0\eta^0$ are shown in Figure 9.
This continuum background is sufficiently large that it must be taken
into account in any analysis of scalar photoproduction, and this is the
primary reason for including it. There are, of course, additional 
continuum background contributions from reactions such as $\gamma p \to 
\Delta(1232)\pi$, $\gamma p \to \Delta(1232)\eta$ and $\gamma p \to 
N(1535) \pi$ that are less amenable to calculation but must also be taken 
into account in analysis.

\subsection{Interference}
\label{interfere}

In any experimental environment, the interference pattern will be
complicated. In addition to interference between a given scalar and the
continuum background there will be interference among the scalars 
themselves. Accordingly we restrict discussion to a simple illustrative 
example.

The formula for the cross section describing the interference between 
direct production of a single scalar, as in Section 3, and the continuum 
background is given in the Appendix. The results for the $f_0(1500)$ in 
scenario I, for constant relative phases of $0^0$, $90^0$, $180^0$ and 
$270^0$ are given in Figure 10.

\bfig[t]
\bc
\begin{minipage}{60mm}
\epsfxsize60mm
\epsffile{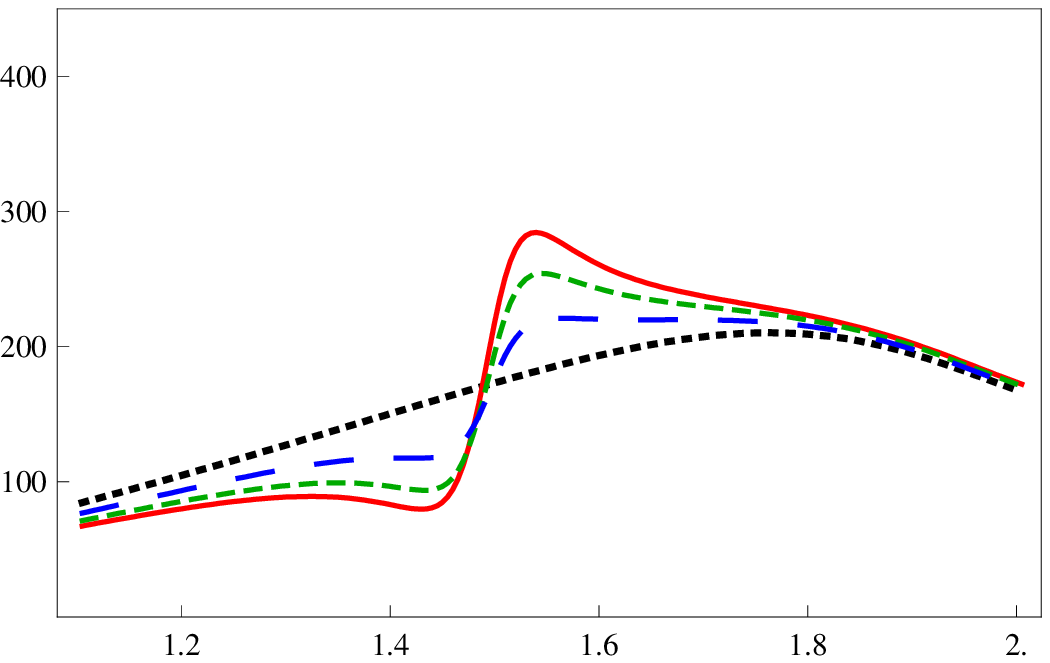}
\begin{picture}(0,0)
\setlength{\unitlength}{1mm}
\put(10,30){{\small $\phi=0^0$}}
\put(-6,36){$\frac{d\sigma}{dM}$}
\put(40,1){{\small$M$ (GeV)}}
\end{picture}
\end{minipage}
\hskip1truecm
\begin{minipage}{60mm}
\epsfxsize60mm
\epsffile{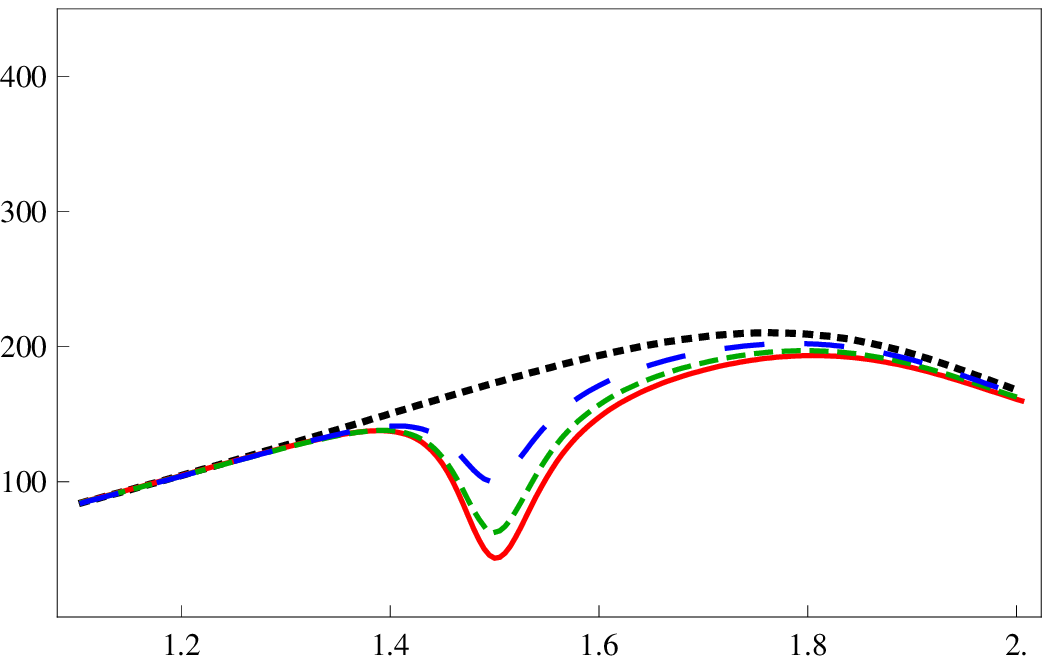}
\begin{picture}(0,0)
\setlength{\unitlength}{1mm}
\put(10,30){{\small $\phi=90^0$}}
\put(-6,36){$\frac{d\sigma}{dM}$}
\put(40,1){{\small$M$ (GeV)}}
\end{picture}
\end{minipage}
\vskip5truemm
\begin{minipage}{60mm}
\epsfxsize60mm
\epsffile{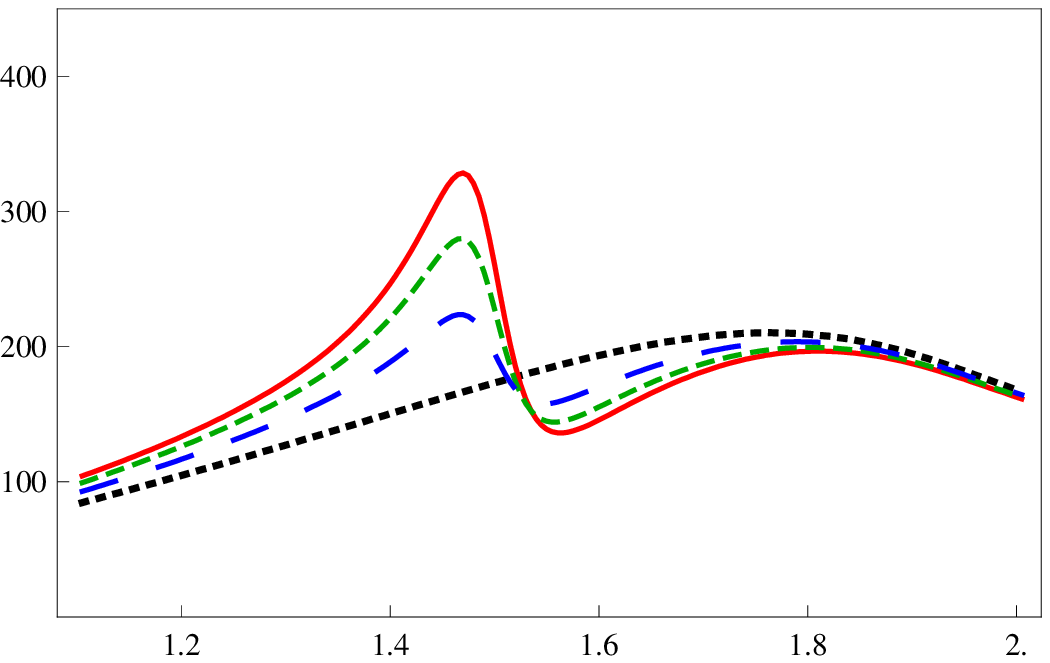}
\begin{picture}(0,0)
\setlength{\unitlength}{1mm}
\put(10,30){{\small $\phi=180^0$}}
\put(-6,36){$\frac{d\sigma}{dM}$}
\put(40,1){{\small$M$ (GeV)}}
\end{picture}
\end{minipage}
\hskip1truecm
\begin{minipage}{60mm}
\epsfxsize60mm
\epsffile{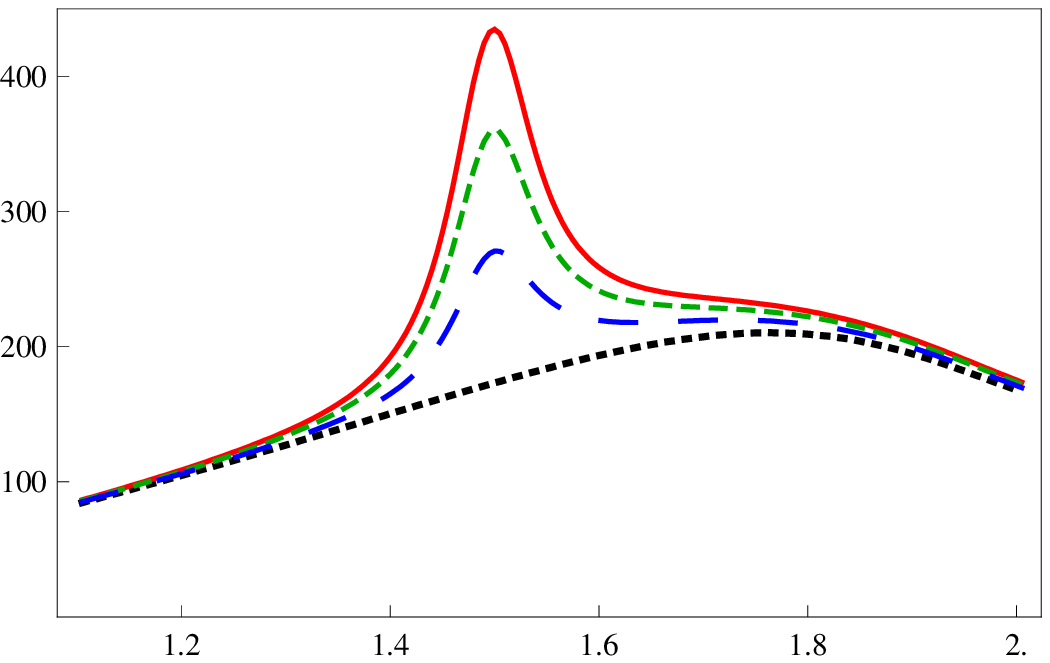}
\begin{picture}(0,0)
\setlength{\unitlength}{1mm}
\put(10,30){{\small $\phi=270^0$}}
\put(-6,36){$\frac{d\sigma}{dM}$}
\put(40,1){{\small$M$ (GeV)}}
\end{picture}
\end{minipage}
\ec
\label{inter}
\caption{Interference cross section in nb GeV$^{-1}$ between the 
$\pi^0\pi^0$
continuum background and the $f_0(1500)$ for different values of the 
relative
phase $\phi$ in scenario I. The glueball masses are L (red), M (green) 
and H
(blue) in each figure.}
\efig

\section{Meson Loop Mechanism}
\label{loops}
 
As was already mentioned in the Introduction, radiative transitions of 
scalars can also proceed via intermediate meson loops. This is widely 
discussed in connection with the possibility of scalar resonances 
generated dynamically. Indeed, if the scalars contain significant 
admixtures of compact quark states, then both quark loops and meson loops 
contribute to the radiative transition amplitude, while in the case of 
dynamically generated resonances (molecules) only meson loops contribute.
 
The most well-studied case is transition via loops of charged 
pseudoscalars, in which case the $\gamma SV$ vertex takes the form 
of equation (\ref{gsv}) with the coupling $g_S(m_S^2,m_V^2)$ given by
\be
g_S(m_S^2,m_V^2)=e\frac{g_{PPV}g_{PPS}}{2\pi^2m_P}I_P(a,b),
\ee
where $a=m_V^2/m_P^2$, $b=m_S^2/m_P^2$, $m_P$ is the mass of the
pseudoscalar in the loop, $g_{PPV}$ and $g_{PPS}$ are the $VP^+P^-$ 
and $SP^+P^-$ coupling constants. The explicit expression for the loop 
integral function $I_P(a,b)$ is given in the Appendix. 

As shown in \cite{KKNHH06}, the decay rates involving $f_0(980)$ and
$a_0(980)$ exhibit a distinct hierarchy pattern: the closer the mass of
the vector meson to the $K \bar K$ threshold, the larger is the
contribution of the kaon loop. So the intermediate kaon-loop mechanism
should dominate the $\phi \to \gamma S$ decay amplitude, as suggested in
\cite{AI89}. The estimates \cite{KKNHH06} for scalar radiative widths in 
the $K \bar
K$ molecular model for scalars are
\be
\Gamma(a_0/f_0 \to \gamma \rho/\omega) \approx 3~{\rm keV},
\label{kaonloopwidth}
\ee
in contrast to the quark-loop results of Table 2.
Thus the decays $a_0/f_0 \to \gamma \rho/\omega$ provide strong 
discrimination between models for these scalars.

In scenario I the $a_0(980)$ and $f_0(980)$ are not $n\bar{n}$ 
states. In this section we consider the extreme possibility of a 
pure molecular assignment for them,
in which case the relevant photoproduction mechanism is via a meson loop.
The calculation of the $a_0(980)$ cross section is straightforward as
only the kaon loop contributes. In the $f_0(980)$ case
there is also a contribution from the $\pi^+\pi^-$ loop. The 
photoproduction formalism is presented in the Appendix, and mass 
distributions are shown in Figure 11. 
The integrated cross sections are $1.3$ nb for the $f_0(980)$ and $0.5$ nb for 
the $a_0(980)$. As expected, these cross sections are small, in 
accordance with general arguments given in \cite{KKNHH06}.

\bfig
\bc
\begin{minipage}{60mm}
\epsfxsize60mm
\epsffile{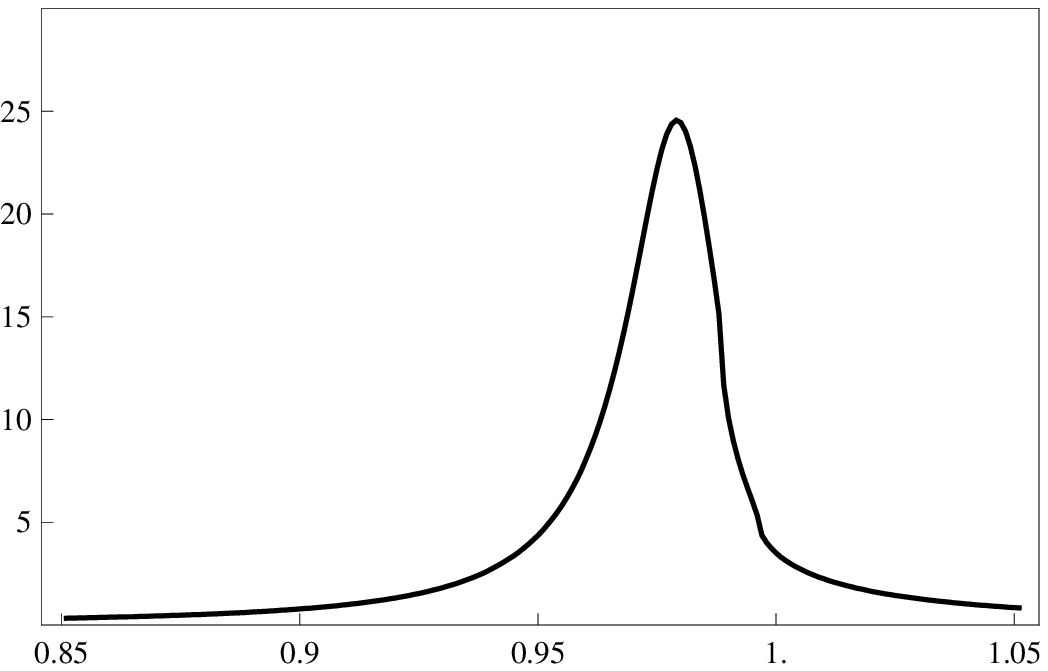}
\begin{picture}(0,0)
\setlength{\unitlength}{1mm}
\put(10,36){{\small (a)}}
\put(-6,36){$\frac{d\sigma}{dM}$}
\put(40,1){{\small $M$ (GeV)}}
\end{picture}
\end{minipage}
\hskip1truecm
\begin{minipage}{60mm}
\epsfxsize60mm
\epsffile{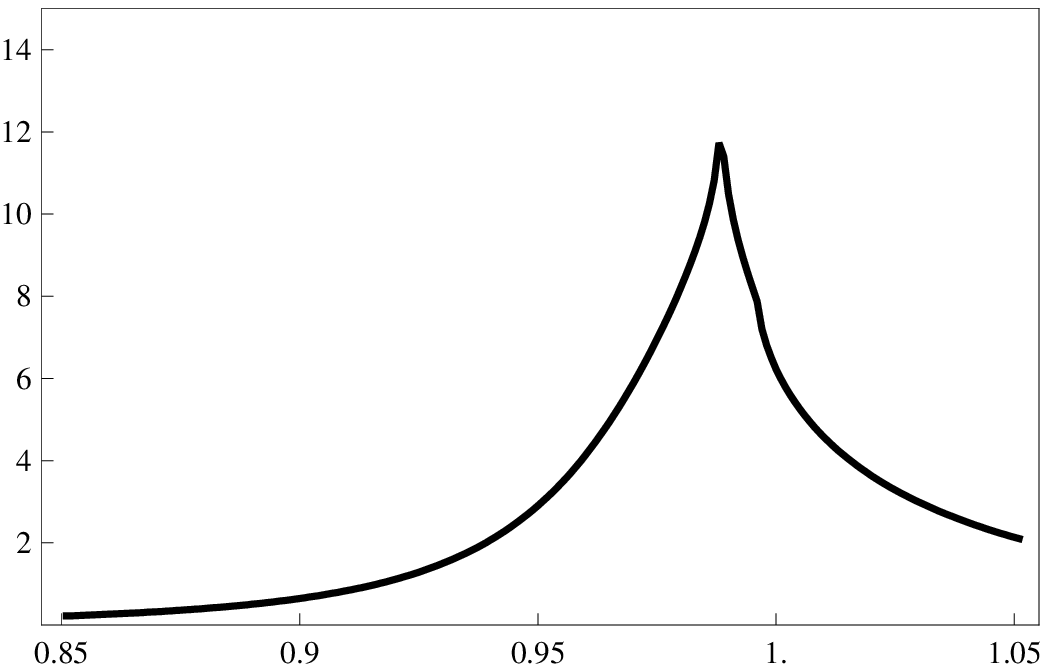}
\begin{picture}(0,0)
\setlength{\unitlength}{1mm}
\put(10,36){{\small (b)}}
\put(-6,36){$\frac{d\sigma}{dM}$}
\put(40,1){{\small $M$ (GeV)}}
\end{picture}
\end{minipage}
\caption{(a) Differential $\pi^0\pi^0$ mass distribution in nb GeV$^{-1}$ 
at $E_{\gamma}=5$ GeV for the $f_0(980)$ in the pseudoscalar loop model.
(b) Differential $\pi^0\eta$ mass distribution in nb GeV$^{-1}$
at $E_{\gamma}=5$ GeV for the $a_0(980)$ in the pseudoscalar loop model.}
\ec
\efig

Recently the contribution of intermediate vector meson channels to  
scalar radiative decays has been considered \cite{NRO08,NROZ08}.
In particular, non-negligible contributions to the $a_0(980) \to \gamma
\omega$ (\cite{NRO08}) and $f_0(1710) \to \gamma \rho$ (\cite{NROZ08}) 
amplitudes were obtained. There is, however, an important difference 
between a purely pseudoscalar loop and one with vector mesons. The former
amplitude is finite (see the discussion in \cite{KKNHH05,KKNHH07}), while 
the latter diverges logarithmically and a cut-off is needed. It is 
claimed in \cite{NRO08,NROZ08} that this divergence can be properly treated,
though the details of the cut-off dependence are not given there.
  
In the photoproduction context, including intermediate 
vector mesons corresponds to taking into account 
final state interactions in the background graphs of Figure 9 (and, also, 
including $K$ and $K^*$ mesons in the loop).  The $\pi^+\pi^-$ and 
$K^+K^-$ \hbox{S-wave} photoproduction has been treated in \cite{Ji98} in such 
an approach. Corresponding contributions are to be taken into account 
in the analysis of the photoproduction of neutral pairs as well, as 
they are potentially important.

\section{Conclusions}
\label{conclusions}

It is clear that light-quark scalar meson photoproduction on protons is a
practical proposition given the luminosities available to modern 
photoproduction. Although we have limited the discussion
to neutral pseudoscalar-pseudoscalar final states, most of the cross
sections we have obtained are sufficiently large to allow for limited
acceptance for these channels. Unfortunately the cross sections on
$^4$He are small for the isoscalars due to their dominant radiative
decay being to $\rho\gamma$. Of course for the isovectors, with their
dominant radiative decay being to $\omega\gamma$, the cross sections
for photoproduction on $^4$He are comparable to those on protons.

To resolve all the issues discussed in the Introduction requires
obtaining the photoproduction cross section for at least two scalars,
but there are exceptions. There are two cases where the difference
between different models for a particular scalar is so great that the
issue can be settled by measuring the cross section for that scalar
alone.

The first of these is the question of the nature of the $a_0(980)$ and
$f_0(980)$, in particular whether there is a significant, or even
dominant, $n\bar{n}$ component in their wave function. The full 
$n\bar{n}$
cross section is given in Table 5 and an estimate of the non-$n\bar{n}$
cross section in Section \ref{loops}. Both the $a_0(980)$ and $f_0(980)$ 
have been observed in photoproduction, at CLAS \cite{JLab2} with quasi-real 
photons from a 5.75 GeV electron beam and at CB-ELSA at $\gamma p$ 
centre-of-mass energies up to 2.55 GeV \cite{UT08,Fuchs}. The CLAS data 
have not been fully analysed, but $\pi^0\eta$ and $\pi^0\pi^0$ mass plots 
(not acceptance corrected) show clear evidence for the $a_0(980)$ and 
$f_0(980)$ respectively. Preliminary results from ELSA also show evidence 
for $a_0(980)$ 
\cite{UT08} and $f_0(980)$ \cite{Fuchs} photoproduction at the upper end 
of their energy ranges. These data point to an admixture of $n\bar{n}$
in the $a_0(980)$ and $f_0(980)$ wave functions. Given the small cross 
section anticipated in the molecular model, even a modest admixture of 
$n\bar{n}$ in the wave function will dominate the cross section.

The $f_0(1370)$ provides the second case where measuring the cross 
section would resolve the issue of its $\pi\pi$ branching fraction and 
possibly also the issue of its existence. The cross section for 
$f_0(1370)$ photoproduction at $E_\gamma = 5$ GeV varies from scenario to 
scenario. In scenarios II and III the $f_0(1370)$ does not exist. In 
scenario I the cross section varies from
27 nb to 94 nb as the glueball mass varies from light to heavy and in 
scenario IV, as the octet member of the ground-state nonet, the cross 
section is 140 nb.
So the non-zero cross sections vary by a factor of 5. The conventional
branching fraction is $2.7\%$, but in the analyses of \cite{Bugg07,AO08} 
it is closer to $80\%$, that is a factor of 30 larger. So for a $\pi\pi$ 
branching fraction of $2.7\%$, the cross section times the $\pi^0\pi^0$ 
branching fraction lies in the range 0.24 to 1.26 nb, while for a 
$\pi\pi$ branching fraction of $80\%$ the range is 7.2 to 37.3 nb. The 
CLAS $\pi^0\pi^0$ data cover this mass range and so, in principle, could be 
used.

For the $f_0(1500)$ the photoproduction cross sections in various 
scenarios are very similar, so that it is not necessarily possible to use 
the $f_0(1500)$ by itself to resolve the ambiguities surrounding the 
nature of the scalars. The cross sections in scenario I with a heavy 
glueball is 17 nb, in scenario II as the octet member of the ground-state 
nonet it is 35 nb and in scenarios III and IV as a member of the first 
radial excitation the cross section is 25 nb,
although this should probably be considered as an upper limit due to its
sensitivity to the wave functions. However in scenario I, if the glueball 
mass is in the light to medium range, with cross sections of 90 nb and 52 
nb respectively, then a clear result can be obtained. Otherwise results 
on the $f_0(1500)$ must be combined with those for the $f_0(980)$ or the 
$f_0(1370)$.

In principle, photoproduction of the $a_0(1450)$ can also provide some
discrimination aince the cross section as a member of the ground-state 
nonet (98 nb) is nearly a factor of 5 larger than that as a member of the 
first radial excitation (21 nb). Unfortunately the considerable 
uncertainty in the branching fractions \cite{PDG06} does not make this 
feasible at present. However some information on $a_0(1450)$ 
photoproduction could, in principle,
be obtained from the CLAS $\pi^0\eta^0$ data as they cover the relevant 
mass range.

Because of the large 4$\pi$ branching fraction of the $f_0(1370)$ (with 
the conventional values) and the $f_0(1500)$, and the implied large 
5$\pi$ branching fraction of the $a_0(1450)$, if measurement of these 
channels were technically feasible then this would not only add to our 
information about the scalars but would assist in resolving their nature.

\section*{Acknowledgements}
We are grateful to Marco Battaglieri, David Bugg, Eberhard Klempt, Wolfgang 
Ochs and Ulrike Thoma for helpful and informative discussions. 

Yu.S.K. acknowledges the support of the Federal Agency for Atomic Energy of 
the Russian Federation, grants RFFI-05-02-04012-NNIOa, DFG-436 
RUS113/820/0-1(R), Nsh-4961.2008.2, and the Federal Programme of the 
Russian Ministry of Industry, Science, and Technology No. 40.052.1.1.1112.

\newpage

\section*{Appendix}

\setcounter{equation}{0}
\renewcommand{\theequation}{A.\arabic{equation}}
{\bf A.1~~Narrow-width cross section}
\label{app1}

\medskip

To obtain the formula (\ref{msquare}) note that we can write
\be
A_{\mu\nu}\gamma_\nu+B_\mu = M_{\mu\nu}(a\gamma_\nu+bp_{1\nu}) =
aM_{\mu\nu}\gamma_\nu+bN_\mu
\ee
with
\be
M_{\mu\nu} = g_{\mu\nu}(q\cdot p)-p_\mu 
q_\nu,~~~~N_\nu=M_{\mu\nu}p_{1\nu}.
\ee

The required trace is
\beqa
T_C &=&{\textstyle \frac{1}{4}}Tr\{(\gamma\cdot p_2 + 
m_p)(aM_{\mu\lambda}
\gamma_\lambda+bN_\mu)(\gamma\cdot p_1 + m_p)(a^*M_{\mu\nu}
\gamma_\nu+b^*N_\mu)\}\nonumber\\
&=&{\textstyle \frac{1}{4}}aa^*M_{\mu\lambda}M_{\mu\nu}Tr\{(\gamma\cdot 
p_2
+m_p)\gamma_\lambda(\gamma\cdot p_1+m_p)\gamma_\nu\}\nonumber\\
&&+{\textstyle \frac{1}{4}}a^*bN_\mu M_{\mu\nu}Tr\{(\gamma\cdot p_2+m_p)
(\gamma\cdot p_1+m_p)\gamma_\nu\}\nonumber\\
&&+{\textstyle \frac{1}{4}}ab^*M_{\mu\lambda}N_\mu Tr\{(\gamma\cdot 
p_2+m_p)
\gamma_\lambda(\gamma\cdot p_1+m_p)\}\nonumber\\
&&+{\textstyle \frac{1}{4}}bb^*N_\mu N_\mu Tr\{(\gamma\cdot p_2+m_p)
(\gamma\cdot p_1+m_p)\}.
\label{trace1a}
\eeqa

The basic traces entering (\ref{trace1a}) are
\beqa
{\textstyle\frac{1}{4}}Tr\{(\gamma\cdot p_2+m_p)\gamma_\lambda
(\gamma\cdot p_1+m_p)\gamma_\nu\} &=& g_{\lambda\nu}(m_p^2-p_1\cdot p_2)
+p_{2\lambda} p_{1\nu}+p_{1\lambda} p_{2\nu}\nonumber\\
{\textstyle\frac{1}{4}}Tr\{(\gamma\cdot p_2+m_p)\gamma_\lambda
(\gamma\cdot p_1+m_p)\} &=& m_p(p_{1\lambda}+p_{2\lambda})\nonumber\\
{\textstyle\frac{1}{4}}Tr\{(\gamma\cdot p_2+m_p)(\gamma\cdot 
p_1+m_p)\}&=&
m_p^2+p_1\cdot p_2
\label{trace1b}
\eeqa

Inserting (\ref{trace1b}) in (\ref{trace1a}) gives
\beqa
T_C &=& aa^*((m_p^2-(p_1\cdot p_2))M_{\mu\nu}M_{\mu\nu}+2N_\mu N_\mu)
\nonumber\\
&&+2m_p(a^*b+ab^*)N_\mu N_\mu+(m_p^2+(p_1\cdot p_2))bb^*N_\mu N_\mu
\label{trace1c}
\eeqa

Equation (\ref{trace1c}) can be rewritten in terms of invariants as
\beqa
&&aa^*\Big({\textstyle\frac{1}{2}}s^2t+{\textstyle\frac{1}{2}}st
(t-2m_p^2-m_S^2)
+{\textstyle\frac{1}{4}}(2m_p^4t+t(t-m_S^2)^2+m_p^2(-2tm_S^2+2m_S^4))\Big)
\nonumber\\
&&+(ab^*+a^*b)\Big(m_ps^2t+m_pst(-2m_p^2+t-m_S^2)+m_p^3(m_p^2t-m_S^2t+m_S^4)
\Big)\nonumber\\
&&+
bb^*(4m_p^2-t){\textstyle\frac{1}{8}}\Big(s^2t-st(2m_p^2-t+m_S^2)+
m_p^2(m_p^2t-m_S^2t+m_S^4)\Big)
\label{trace3a}
\eeqa

The result (\ref{trace3a}) can be written compactly in terms of the
kinematical bounaries $t_1$ and $t_2$ which are given by
\beqa
t_{1,2} &=& \frac{1}{2s}\Big(-(m_p^2-s)^2+m_S^2(m_p^2+s)\nonumber\\
&&\pm (m_p^2-s)\surd((m_p^2-s)^2-2m_S^2(m_p^2+s)+m_S^4)\Big),
\label{t12a}
\eeqa
so that for the exchange of a single vector meson
\beqa
|M(s,t)|^2=-T_C &=& -\textstyle{\frac{1}{2}}aa^*(s(t-t_1)(t-t_2)+
\textstyle{\frac{1}{2}}st(t-m_S^2)^2)\nonumber\\
&&-\textstyle{\frac{1}{2}}(ab^*+a^*b)m_ps(t-t_1)(t-t_2)\nonumber\\
&&-\textstyle{\frac{1}{8}}bb^*s(4m_p^2-t)(t-t_1)(t-t_2).
\label{msquarea}
\eeqa
Finally the differential cross section is given by
\be
\frac{d\sigma}{dt}=-\frac{T_C}{16\pi(s-m_p^2)^2}.
\label{sigmaa}
\ee
Obviously in practice the amplitudes for $\rho$ and $\omega$ exchange are
added coherently.

\bigskip

{\bf A.2~~Signal cross section}
\label{app2}
\medskip

The differential cross section for the production of a scalar, mass $M$, and 
its decay to two pseudoscalars, masses $m_a$ and $m_b$, is
\be
\frac{d\sigma_S}{dt~dM~d\Omega}=-\frac{1}{2^8\pi^4}\frac{k_i(M,m_a,m_b)}
{(s-m_p^2)^2}T_S.
\ee
with $k_i(M,m_a,m_b)=\sqrt{(M^2-(m_a+m_b)^2)(M^2-(m_a-m_b)^2)}/2M$.

Following (\ref{trace1c}), $T_S$ is 
\beqa
T_S &=& a_Sa^*_S((m_p^2 - (p_1\cdot p_2))M_{\mu\nu}M_{\mu\nu}+2N_\mu N_\mu)
\nonumber\\
&&+2m_p(a_Sb^*_S+b_Sa^*_S)N_\mu N_\mu+(m_p^2+(p_1\cdot p_2))b_Sb^*_S
N_\mu N_\mu,
\label{app2_1}
\eeqa
with
\beqa
a_S &=& \frac{g_i}{m_S^2 - M^2 -iM\Gamma_{\rm Tot}}
(g_{S\rho}(g_{V\rho}+2m_pg_{T\rho})D_\rho+
g_{S\omega}(g_{V\omega}-2m_pg_{T\omega})D_\omega)\nonumber\\
b_S &=& -\frac{g_i}{m_S^2 - M^2 -iM\Gamma_{\rm Tot}}
(2g_{S\rho}g_{T\rho}D_\rho+2g_{S\omega}g_{T\omega}D_\omega).
\label{app2_3}
\eeqa
Both $\rho$ and $\omega$ exchanges have been included explicitly and
$D_\rho$, $D_\omega$ are the Regge propagators. .

The decay constant $g_i$ in (\ref{app2_3}) is defined in terms of the partial 
width $\Gamma_i$ at resonance by
\be
\Gamma_i = \frac{g_i^2\rho_i(m_S,m_a,m_b)}{16\pi m_S}
\label{gdef}
\ee
with 
\be
\rho_i(M,m_a,m_b)=\sqrt(1-(m_a+m_b)^2/M^2)(1-(m_a-m_b)^2)/M^2) = 
2k_i(M,m_a,m_b)/M.
\label{rhodef}
\ee

Substituting (\ref{gdef}) in (\ref{app2_3}) and recalling (\ref{sigmaa})
gives 
\be
\frac{d\sigma}{dt~dM}=\frac{d\sigma_0(t,M)}{dt}\frac{2m_S^2}{\pi}
\frac{\Gamma_i(M)}{(m_S^2-M^2)^2+M^2\Gamma_{\rm Tot}^2},
\label{signaldcsa}
\ee
where $d\sigma_0(t,M)/dt$ is the narrow-width cross section at the scalar mass 
$M$.
\newpage

{\bf A.3~~Background cross section}
\label{app3}
\medskip

First note the simplification of (\ref{backamp1})
\beqa
\epsilon_{\mu\rho\beta\gamma}q_{\beta}v_{\gamma}
\epsilon_{\rho\nu\lambda\sigma}p_{\lambda}v_{\sigma}&=&-g_{\mu\nu}
\Big((q\cdot p)v^2 - (q\cdot v)(v.\cdot p)\Big)-v_\mu\Big(q_\nu(p\cdot v)
-v_\nu(q\cdot p)\Big)
\nonumber\\
&&-p_\mu\Big(v_\nu(q\cdot v) - q_\nu v^2\Big)
\eeqa
and define, with $i=1,2$,
and define, with $i=1,2$,
\beqa
M_{\mu\nu}^{(i)} &=& g_{\mu\nu}b_i+v_{i\mu}c_{i\nu}+p_\mu 
d_{i\nu}\nonumber\\
b_i &=& (q\cdot p)^2 v_i^2 - (q\cdot v_i)(p\cdot v_i)\nonumber\\
c_{i,\nu} &=& q_{\nu}(p\cdot v_i) - v_{i\nu}(q\cdot p)\nonumber\\
d_{i\nu} &=& v_{i\nu}(q\cdot v_i)-q_\nu v_{i}^2,
\eeqa
where $v_i=k_i-q$, with $k_1$ and $k_2$ the 4-momenta of the 
pseudoscalars.

The current can be written as
\be
M_\mu = M_{\mu\nu}^{(1)}(\alpha_1\gamma_\nu+\beta_1p_{1\nu})+
M_{\mu\nu}^{(2)}(\alpha_2\gamma_\nu+\beta_2p_{1\nu}).
\ee
It is convenient to define
\be
C_{\pi\pi} = 1~~~~C_{\eta\eta} = {\textstyle\frac{2}{3}}~~~~C_{\pi\eta} =
{\textstyle\sqrt{\frac{2}{3}}}
\label{cdef}
\ee
Then, for the $\pi^0\pi^0$ final state,
\beqa
\alpha_1 &=& C_{\pi\pi}[(g_{\rho V}+2m_pg_{\rho 
T})D_{\rho}\Pi_{\omega}(v_1)
+{\textstyle\frac{1}{3}}(g_{\omega V}+2m_pg_{\omega T})D_{\omega}
\Pi_{\rho}(v_1)]
\nonumber\\
\beta_1 &=& -2C_{\pi\pi}[g_{\rho T}D_{\rho}\Pi_{\omega}(v_1)+
{\textstyle\frac{1}{3}}g_{\omega T}D_{\omega}\Pi_{\rho}(v_1)]
\nonumber\\
\alpha_2 &=& C_{\pi\pi}[(g_{\rho V}+2m_pg_{\rho 
T})D_{\rho}\Pi_{\omega}(v_2)
+{\textstyle\frac{1}{3}}(g_{\omega V}+2m_pg_{\omega T})D_{\omega}
\Pi_{\rho}(v_2)]
\nonumber\\
\beta_2 &=&-2C_{\pi\pi}[g_{\rho T}D_{\rho}\Pi_{\omega}(v_2)+
{\textstyle\frac{1}{3}}g_{\omega T}D_{\omega}\Pi_{\rho}(v_2)],
\eeqa
for $\eta^0\eta^0$,
\beqa
\alpha_1 &=& C_{\eta\eta}[(g_{\rho V}+2m_pg_{\rho 
T})D_{\rho}\Pi_{\rho}(v_1)
+{\textstyle\frac{1}{3}}(g_{\omega V}+2m_pg_{\omega T})D_{\omega}
\Pi_{\omega}(v_1)]
\nonumber\\
\beta_1 &=& -2C_{\eta\eta}[g_{\rho T}D_{\rho}\Pi_{\rho}(v_1)+
{\textstyle\frac{1}{3}}g_{\omega T}D_{\omega}\Pi_{\omega}(v_1)]
\nonumber\\
\alpha_2 &=& C_{\eta\eta}[(g_{\rho V}+2m_pg_{\rho 
T})D_{\rho}\Pi_{\rho}(v_2)
+{\textstyle\frac{1}{3}}(g_{\omega V}+2m_pg_{\omega T})D_{\omega}
\Pi_{\omega}(v_2)]
\nonumber\\
\beta_2 &=& -2C_{\eta\eta}[g_{\rho T}D_{\rho}\Pi_{\rho}(v_2)
+{\textstyle\frac{1}{3}}g_{\omega T}D_{\omega}\Pi_{\omega}(v_2)]
\eeqa
and for $\pi^0\eta^0$
\beqa
\alpha_1 &=& C_{\pi\eta}[{\textstyle\frac{1}{3}}(g_{\rho V}+2m_pg_{\rho 
T})D_{\rho}\Pi_{\rho}(v_1)
+(g_{\omega V}+2m_pg_{\omega T})D_{\omega}
\Pi_{\omega}(v_1)]
\nonumber\\
\beta_1 &=& -2C_{\pi\eta}[{\textstyle\frac{1}{3}}g_{\rho T}D_{\rho}\Pi_{\rho}(v_1)+
g_{\omega T}D_{\omega}\Pi_{\omega}(v_1)]
\nonumber\\
\alpha_2 &=& C_{\pi\eta}[{\textstyle\frac{1}{3}}(g_{\rho V}+2m_pg_{\rho 
T})D_{\rho}\Pi_{\omega}(v_2)
+ (g_{\omega V}+2m_pg_{\omega T})
D_{\omega}\Pi_{\rho}(v_2)]
\nonumber\\
\beta_2 &=& -2C_{\pi\eta}[({\textstyle\frac{1}{3}}g_{\rho T}D_{\rho}\Pi_{\omega}(v_2)+
g_{\omega T}D_{\omega}
\Pi_{\rho}(v_2))].
\eeqa

As $M_\mu$ has the structure
\be
M_\mu = A_{\mu\lambda}\gamma_\lambda + B_\mu
\label{btrace1a}
\ee
we have, as before, to calculate the trace
\be
T_B = {\textstyle \frac{1}{4}}Tr\{(\gamma\cdot p_2 + m_p)
(A_{\mu\lambda}\gamma_\lambda + B_\mu)(\gamma\cdot p_1 + m_p)
(A_{\mu\lambda}\gamma_\lambda + B_\mu)\}.
\label{btrace2a}
\ee
The result is
\beqa
&&T_B =\nonumber\\
&&(m_p^2-(p_1\cdot 
p_2))\Big(\alpha_1\alpha_1^*M^{(1)}_{\mu\nu}M^{(1)}_{\mu\nu}
+(\alpha_1\alpha_2^*+\alpha_2\alpha_1^*)M^{(1)}_{\mu\nu}M^{(2)}_{\mu\nu}
+\alpha_2\alpha_2^*M^{(2)}_{\mu\nu}M^{(2)}_{\mu\nu}\Big)\nonumber\\
&&+2\Big(\alpha_1\alpha_1^*N_1^2+(\alpha_1\alpha_2^*+\alpha_2\alpha_1^*)
(N_1\cdot N_2)+\alpha_2\alpha_2^*N_2^2\Big)\nonumber\\
&&+2m_p\Big((\beta_1\alpha_1^*+\beta_1^*\alpha_1)N_1^2+(\beta_1\alpha_2^*
+\alpha_1^*\beta_2+\beta_1^*\alpha_2+\beta_2^*\alpha_1)(N_1\cdot N_2)
\nonumber\\
&&+(\beta_2\alpha_2^*+\beta_2^*\alpha_2)N_2^2\Big)\nonumber\\
&&+(m_p^2+(p_1\cdot p_2))\Big(\beta_1\beta_1^*N_1^2+(\beta_1\beta_2^*+
\beta_1^*\beta_2)(N_1\cdot N_2)+\beta_2\beta_2^*N_2^2\Big),
\label{btrace3a}
\eeqa
with
\be
N_{i\mu} = p_{1\mu}b_i+v_{i\mu}(p_1\cdot c_i)+p_\mu(p_1\cdot d_i).
\label{btrace4a}
\ee
The background cross section is then given by
\be
\frac{d\sigma_B}{dt~dM~d\Omega} = -\zeta^2 \frac{g_0^2}{2^8\pi^4}
\frac{k}{(s-m_p^2)^2}T_B.
\label{bsigmaa}
\ee
and the factor $\zeta = 1$ for different pseudoscalars and 
$\frac{1}{\sqrt{2}}$
for identical pseudoscalars.

\bigskip

{\bf A.4~~Interference cross section}
\label{app4}
\medskip

We assume a constant phase $\phi$ between the continuum background and 
the
resonance signal. The required trace for the interference term is
\beqa
T_{int} &=& (m_p^2-(p_1\cdot 
p_2))(\alpha_1a^*_SM^{(1)}_{\mu\nu}M_{\mu\nu}
+\alpha_2a^*_SM^{(2)}_{\mu\nu}M_{\mu\nu}+\nonumber\\
&&2(\alpha_1a^*(N^{(1)}\cdot N)+\alpha_2a^*_S(N^{(2)}\cdot N)\nonumber\\
&&2m_p((\beta_1a^*_S+\alpha_1b^*_S)(N^{(1)}\cdot N))+
(\beta_2a^*_S+\alpha_2b^*_S)(N^{(2)}\cdot N))+\nonumber\\
&&(m_p^2+(p_1\cdot p_2))(\beta_1b^*_S(N^{(1)}\cdot N)+
\beta_2b^*_S(N^{(2)}\cdot N))
\label{app4_1}
\eeqa
The interference cross section is then
\be
\frac{d\sigma^{int}}{dt~dM~d\Omega} = -\zeta
\frac{1}{2^8\pi^4}\frac{g_0k}{(s-m_p^2)^2}(\cos\phi(T_{int}+T^*_{int})
+i\sin\phi(T_{int}-T^*_{int})).
\label{app4_2}
\ee

\newpage

{\bf A.5~~Meson-loop cross section}
\label{app5}
\medskip

The loop integral function $I_P(a,b)$ is
(see e.g. \cite{AI89}, \cite{CIK93}, and, for $x<0$, \cite{oset})
\be
I_P(a,b)=\frac{1}{2(a-b)}-\frac{2}{(a-b)^2}\left[f(\frac{1}{b})-
f(\frac{1}{a})\right]+\frac{a}{(a-b)^2}\left[g(\frac{1}{b})-
g(\frac{1}{b})\right],
\label{loopintegral}
\ee
\be
f(x)=\left\{
\begin{array}{ll}
-\left[\arcsin(\frac{1}{2\sqrt{x}})\right]^2,&x>\frac{1}{4}\\
\frac{1}{4}\left[\ln(\frac{\eta_+}{\eta_-})-i\pi\right]^2,&0<x<\frac{1}{4}\\
\left[\ln(\frac{1}{2\sqrt{-x}}+\sqrt{1-\frac{1}{4x}})\right]^2,&x<0
\end{array}
\right.
\ee
\be
g(x)=\left\{
\begin{array}{ll}
\sqrt{4x-1}\arcsin(\frac{1}{2\sqrt{x}}),&x>\frac{1}{4}\\
\frac{1}{2}\sqrt{1-4x}\left[\ln(\frac{\eta_+}{\eta_-})-i\pi\right],
&0<x<\frac{1}{4}\\
\sqrt{1-4x}\ln(\frac{1}{2\sqrt{-x}}+\sqrt{1-\frac{1}{4x}}),&x<0
\end{array}
\right.
\ee
and
\be
\eta_{\pm}=\frac{1}{2x}\left[1\pm\sqrt{1-4x}\right].
\ee

For a given two-meson final state
$ab$
\be
\frac{d\sigma(ab)}{dtdMd\Omega}=-\frac{1}{2^8\pi^4}\frac{k}{(s-m_p^2)^2}
T_{l,ab},
\ee
where the trace $T_{l,ab}$ is given by
\beqa
T_{l,ab} &=& |A_{l,ab}|^2(m_p^2-(p_1 \cdot p_2))M_{\mu\nu}M_{\mu\nu}
\nonumber\\
&&+N_{\mu}N_{\mu}[2|A_{l,ab}|^2+2m_p(A_{l,ab}B^*_{l,ab}+A_{l,ab}^*B_{l,ab})
\nonumber\\
&&+(m_p^2+(p_1 \cdot p_2)|B_{l,ab}|^2].
\label{traceloop}
\eeqa
The quantities $A_{l,ab}$ and $B_{l,ab}$ are
\be
A_{l,ab}=A_{l,ab}^{\rho}+A_{l,ab}^{\omega},
\ee
\be
B_{l,ab}=B_{l,ab}^{\rho}+B_{l,ab}^{\omega},
\ee
\be
A_{l,ab}^{\rho(\omega)}=(g_{\rho(\omega) V}+2m_pg_{\rho(\omega)
T})D_{\rho(\omega)}\sum_P L_{P,\rho(\omega)}
t_{PP \to ab},
\ee
\be
B_{l,ab}^{\rho(\omega)}=-2g_{\rho(\omega) T}D_{\rho(\omega)}\sum_P
L_{P,\rho(\omega)} t_{PP \to ab}.
\ee
Here $D_{\rho}$ and $D_{\omega}$ are the Regge propagators (\ref{regge}),
the sum is over all possible pseudoscalars $P$ in the loop,
\be
L_{P,\rho(\omega)}=
\frac{eg_{\rho(\omega)PP}I_P(t/m_P^2,M^2/m_P^2)}{2\pi^2m_P^2},
\ee
where $m_P$ is the mass of the pseudoscalars, $g_{\rho(\omega)PP}$ is the
$\rho(\omega)PP$ coupling constant, $I_P$ is the loop integral function
(\ref{loopintegral}) and $t_{PP \to ab}$ is the $t$-matrix for the
transition
$PP \to ab$.

In the calculation of the $a_0(980)$ photoproduction cross section
only the kaon loop contributes. The
$\rho(\omega)K^+K^-$ coupling constants can be estimated from that for
the $\rho \to \pi\pi$ decay width with the help of $SU(3)$ symmetry
considerations:
\be
g_{\rho K^+K^-}=g_{\omega K^+K^-}=\textstyle{\frac{1}{2}}g_{\rho \pi
\pi}=2.13.
\ee
For the $t$-matrix we use the parametrisation introduced in \cite{ADS80}:
\be
t_{K^+K^- \to \pi^0\eta}(M)=\frac{g_{a_0\pi\eta}g_{a_0K^+K^-}}{D_{a_0}},
\label{pieta}
\ee
where
\be
D_{a_0}(M)=M_{a_0}^2-M^2+\sum_{ab}(Re\Pi_{ab}(M_{a_0})-\Pi_{ab}(M)),
\label{D}
\ee
\beqa
\Pi_{ab}(M) &=& \frac{g_{a_0ab}^2}{16\pi^2}\Big(-\frac{m_+m_-}{M^2}
\ln\frac{m_a}{m_b} \nonumber\\
&&+\rho_{ab}(M)(i\pi+\ln\frac{\sqrt{M^2-m_-^2}-\sqrt{M^2-m_+^2}}
{\sqrt{M^2-m_-^2}+\sqrt{M^2-m_+^2}})\Big),
\eeqa
where $m_+=m_a+m_b$, $m_-=m_a-m_b$. This expression is valid above
threshold ($m_a+m_b<M$); below threshold one should use the analytical
continuation. In the case of $a_0$, $ab=\pi^0\eta$, $K^+K^-$, $K^0\bar K^0$.

This parametrisation was recently employed in the analysis \cite{KLOE07}
of $\phi \to \pi^0 \eta \gamma$ radiative decays. We use the
$a_0(980)$ parameters found in that analysis (``kaon loop'' version of the 
fit):
\be
M_{a_0}=983~{\rm MeV},~~g_{a_0\pi\eta}=2.8~{\rm GeV},~~g_{a_0K^+K^-}=
g_{a_0K^0\bar K^0}=2.16~{\rm GeV}.
\label{pietaparameters}
\ee

The situation with the $f_0(980)$ is more complicated, as the 
$\pi^+\pi^-$ loop also contributes. For the $t$-matrix we use here the
parametrisation
\be
t_{\pi^+\pi^-\to\pi^0\pi^0}=
\frac{g_{f_0\pi^+\pi^-}g_{\pi^0\pi^0}}{D_{f_0}},~~~~
t_{K^+K^- \to
\pi^0\pi^0}=
\frac{g_{f_0\pi^0\pi^0}g_{f_0K^+K^-}}{D_{f_0}}
\ee 
with $D_{f_0}$ given by
an expression similar to (\ref{D}) (with
$ab=\pi^0\pi^0$, $\pi^+\pi^-$, $K^+K^-$, $K^0\bar K^0$),
and $g_{f_0\pi^+\pi^-}^2=\frac{2}{3}g_{f_0\pi\pi}^2$,
$g_{f_0\pi^0\pi^0}^2=\frac{1}{3}g_{f_0\pi\pi}^2$.

The resonance parameters are taken from the ``kaon loop'' version of the 
fit \cite{KLOE07f} obtained in the analysis of $\phi \to \pi^0\pi^0 
\gamma$ radiative decay:
\be
M_{a_0}=976.8~{\rm MeV},~~g_{f_0\pi^+\pi^-}=-1.43~{\rm 
GeV},~~g_{f_0K^+K^-}=
g_{f_0K^0\bar K^0}=3.76~{\rm GeV}.
\ee
In practice, there is strong interference between the $f_0(980)$
resonance and the \hbox{$S$-wave} isosinglet $\pi\pi$ background, which 
should be taken into account.


\begin{thebibliography}{99}

\bibitem{PDG06}
Particle Data Group, J.Phys. G33 (2006) 1

\bibitem{AC95}
C Amsler and F E Close, Phys.Lett. B353 (1995) 385,
Phys.Rev. D53 (1996) 295

\bibitem{MP99}
C J Morningstar and M Peardon, Phys.Rev. D60 (1999) 034509

\bibitem{MT05}
H B Meyer and M J Teper, Phys.Lett. B605 (2005) 344

\bibitem{Chen06}
Y Chen {\it et al}, Phys.Rev. D73 (2006) 014516

\bibitem{LW00}
W Lee and D Weingarten, Phys.Rev. D61 (2000) 014015

\bibitem{CK01}
F E Close and A Kirk, Eur.Phys.J C21 (2001) 531

\bibitem{LQCD}
C McNeile, arXiv:0710.0985[hep-lat], arXiv:0710.2470[hep-lat]

\bibitem{KZ07}
E Klempt and A Zaitsev, arXiv:0708.4016[hep-ph]

\bibitem{Ochs06}
W Ochs, arXiv:hep-ph/0609207

\bibitem{MO99}
P Minkowski and W Ochs, Eur.Phys.J C9 (1999) 283

\bibitem{Bugg07}
D V Bugg, Eur.Phys.J C52 (2007) 55

\bibitem{AO08}
M Albaladejo and J A Oller, arXiv:0801.4929[hep-ph]

\bibitem{Nar06}
S Narison, Phys.Rev. D73 (2006) 114024

\bibitem{CDK02}
F E Close, A Donnachie and Yu S Kalashnikova, Phys.Rev. D65 (2002) 092003

\bibitem{CDK03}
F E Close, A Donnachie and Yu S Kalashnikova, Phys.Rev. D67 (2003) 074031

\bibitem{GLV97}
M Guidal, J-M Laget and M Vanderhaeghen, Nucl.Phys. A627 (1997) 645

\bibitem{JLab}
G Asryan {\it et al}, JLab experiment E-07-009

\bibitem{KKNHH06}
Yu S Kalashnikova, A Kudryavtsev, A V Nefediev, J Haidenbauer and C 
Hanhart, Phys.Rev. C73 (2006) 045203

\bibitem{RHE87}
R Machleidt, K Holinde and Ch Elster, Phys.Rep. 149 (1987) 1

\bibitem{OO78}
M G Olsson and E T Osypowski, Phys.Rev. D17 (1978) 174

\bibitem{NBL90}
S Nozawa, B Blankenbecler and T-S H Lee, Nucl.Phys. A513 (1990) 459

\bibitem{GG94}
H Garcilazo and E Moya de Guerra, Nucl.Phys. A562 (1994) 521

\bibitem{Wor72}
R P Worden, Nucl.Phys. B37 (1972) 253

\bibitem{BDS74}
I S Barker, A Donnachie and J K Storrow, Nucl.Phys. B79 (1974) 431

\bibitem{BS78}
I S Barker and J K Storrow, Nucl.Phys. B137( 1978) 413

\bibitem{MS91}
H Morita and T Suzuki, Prog.Theor.Phys. 86 (1991) 671

\bibitem{GS78}
G Grammar and D Sullivan in {\it Electromagnetic Interactions of Hadrons}
volume 1, A Donnachie and G Shaw eds, Plenum Press 1978

\bibitem{JLab2}
G Asryan {\it et al}, JLab proposal R-07-009

\bibitem{KLOE07f} 
F Ambrosino {\it et al}, KLOE Collaboration, Eur.Phys.J. C49 (2007) 473

\bibitem{KLOE07} 
F. Ambrosino {\it et al}, KLOE collaboration, contributed
paper to Lepton Photon 2007, arXiv:0707.4609[hep-ex]

\bibitem{WA102}
D Barberis {\it et al}, WA102 Collaboration, Phys.Lett. B479 (2000) 59

\bibitem{CB01}
A Abele {\it et al}, Crystal Barrel Collaboration, Eur.Phys.J. C19 (2001) 
667

\bibitem{CB03}
C A Baker {\it et al}, Phys.Lett. B563 (2003) 140

\bibitem{Bramon92}
A Bramon {\it et al}, Phys.Lett. B283 (1992) 416

\bibitem{AI89}
N N Achasov and V N Ivanchenko, Nucl.Phys. B315 (1989) 465

\bibitem{NRO08} 
H Nagahiro, L Roca and E Oset, arXiv:0802.0455[hep-ph]

\bibitem{NROZ08} 
H Nagahiro, L Roca, E Oset and B S Zou, arXiv:0803.4460[hep-ph]

\bibitem{KKNHH05} 
Yu S Kalashnikova, A E Kudryavtsev, A V Nefediev, C Hanhart and 
J Haidenbauer, Eur.Phys.J. A24 (2005) 437

\bibitem{KKNHH07} 
Yu S Kalashnikova, A E Kudryavtsev, A V Nefediev, C Hanhart and 
J Haidenbauer, arXiv:0711.2902[hep-ph]
 
\bibitem{Ji98}
C-R Ji, R Kaminski, L Lesniak, A Szczepaniak and R Williams, Phys.Rev.
C58 (1998) 1205

\bibitem{UT08}
U Thoma, private communication; I Horn {\it et al}, in preparation

\bibitem{Fuchs}
M Fuchs, Ph.D Thesis, Bonn University

\bibitem{CIK93}
F E Close, N Isgur and S Kumano, Nucl.Phys. B389 (1993) 513

\bibitem{oset}
E Marco, E Oset and H Toki, Phys.Rev. C60 (1999) 015202

\bibitem{ADS80}
N N Achasov, S A Devyanin, and G N Shestakov, Yad.Fiz. 32 (1980) 1098
[Sov.J.Nucl. Phys. 32 (1980) 566]

\end{thebibliography}
\end{document}